\documentclass{iopart}

\usepackage[english]{babel}
\usepackage[T1]{fontenc}
\usepackage{graphicx}
\usepackage{xcolor}
\usepackage[colorlinks]{hyperref}
\usepackage{multirow}
\usepackage{booktabs}
\usepackage[font=small,labelfont=bf,
  justification=justified,
  format=plain]{caption}
\usepackage{subcaption}
\usepackage{multirow}

\begin{document}

\title{Bianchi type-I cosmological dynamics in \texorpdfstring{$f(\mathcal{Q})$}{} gravity: a covariant approach}
\author{Fabrizio Esposito$^{1}$\footnote{fabrizio.esposito01@edu.unige.it}, Sante Carloni$^{1}$\footnote{sante.carloni@unige.it} and Stefano Vignolo$^{1}$\footnote{stefano.vignolo@unige.it}}
\address{$^{1}$ DIME, Università di Genova, Via all'Opera Pia 15, 16145 Genova, Italy}

\begin{abstract}
Making use of the $1 + 3$ covariant formalism, we show explicitly the effect that nonmetricity has on the dynamics of the universe. Then, using the Dynamical System Approach, we analyze the evolution of Bianchi type-I cosmologies within the framework of $f(\mathcal{Q})$ gravity. We consider several models of function $f(\mathcal{Q})$, each of them manifesting isotropic eras of the universe, whether transitional or not. 
In one case, in addition to the qualitative analysis provided by the dynamical system method, we also obtain analytical solutions in terms of the average length scale $l$.
\end{abstract}

\maketitle

\section{Introduction}

In the early twentieth century, Einstein formulated the first geometric theory of gravitation, the ``General Relativity'' (GR) \cite{Einstein:1916vd,Wald:1984rg}. In GR the gravitational field is described in terms of the metric tensor $g_{ij}$ and the spacetime presents a curvature connected to the matter distribution. Curvature and metric are related to each other by the Levi-Civita connection. Right after the formulation of GR, alternative ways to geometrize gravity were explored, for instance leading  Weyl to develop a theory with a symmetrical and nonmetric connection, which aimed to unify gravity and electromagnetism \cite{Weyl:1918ib}. The adjective nonmetric indicates that the metric tensor $g_{ij}$ is not preserved under parallel transport, implying that the inner product between vectors changes when they are transported along a given curve. From the geometric point of view, all this is expressed by the nonmetricity tensor $Q_{hij}$. Another interesting geometric approach was considered  by Cartan, who developed a theory where the connection is not symmetric but metric compatible \cite{E1922}, thus introducing torsional degrees of freedom.

Both nonmetricity and torsion embody aspects of the connection different from the curvature, which, instead, is the only cornerstone of GR.

Over the years, major developments have been made in extensions and generalizations of GR (see e.g. \cite{CAPOZZIELLO2011167}), involving also torsion \cite{Kr_k_2019, Cai:2015emx}. However,
the nonmetric theories came  into the limelight only in $1999$ when the so-called ``Symmetric Teleparallel Gravity'' (STG) \cite{Nester:1998mp,Adak:2005cd,Conroy:2017yln,Adak:2008gd} was proposed. In this theory, gravitation is strictly connected to the nonmetricity tensor and the related nonmetricity scalar $\mathcal{Q}$, while both curvature and torsion are set to zero. A generalization of STG, which recently has gained great attention, is $f(\mathcal{Q})$ gravity \cite{BeltranJimenez:2017tkd}, where the action of the gravitational field is described by a generic function of the nonmetricity scalar.
 
$f(\mathcal{Q})$ gravity has recently attracted great interest. In particular, its cosmology has been studied in detail \cite{Jim_nez_2020,Vignolo:2021frk,Esposito:2021ect, Atayde:2021pgb,Albuquerque:2022eac,Narawade:2022jeg,Lu:2019hra,Khyllep:2021pcu,Dimakis:2021gby,Mandal:2021wer,Hohmann:2021ast,De:2022shr,Barros_2020,Anagnostopoulos:2021ydo,Anagnostopoulos:2022gej,Ayuso:2021vtj,Xu:2019sbp, Xu:2020yeg,Bhattacharjee_2020}, some spherically symmetric models in \cite{Wang:2021zaz,Lin:2021uqa,DAmbrosio:2021zpm,Heisenberg:2022nvs,Mandal:2021qhx}, and wormhole solutions in \cite{Mustafa:2021bfs,Mustafa:2021ykn,Banerjee:2021mqk}.

In the following, we focus on the Bianchi type-I universes in the context of $f(\mathcal{Q})$ theory. Bianchi type-I metrics are the simplest anisotropic generalization of the spatially flat Friedmann-Lema\^{\i}tre-Robertson-Walker (FLRW) cosmologies. We specifically aim to understand how nonmetricity could contribute to driving the actual Universe to be essentially isotropic.

For our study, we endow the spacetime with a congruence of time-like curves, whose tangent vector field $u$ determines, at each point, the local direction of the time flow.  The existence of this vector field implies the existence of preferred rest frames at each point. Thus, the field $u$ can be thought as the $4$-velocity field of a family of observers whose world lines coincide with the congruence. This assumption is the basis of the so-called $1+3$ covariant approach \cite{ellis_maartens_maccallum_2012,Roy:2014lda,poisson_2004}. The reason behind the use of this formalism is that it gives us a direct insight into the physical relevance of nonmetricity in both geometric and dynamical aspects of the spacetime. Specifically, we will see how nonmetricity affects the expansion and anisotropy in Bianchi type-I cosmologies. Other $1+3$ approaches to $f(\mathcal{Q})$ cosmology are given in \cite{Iosifidis:2018diy, Yang:2021fjy}.

Since the resulting cosmological equations are in general difficult to solve, we attempt to obtain a global picture of the cosmic evolution by means of the so-called Dynamical Systems Approach (DSA), see e.g. \cite{perko2012differential, Bahamonde:2017ize}. This technique allows us to study a cosmological model by analyzing the behavior of the orbits in a phase space connected with the geometrical features and matter sources of space-time.  Making use of DSA, it is possible to achieve a semi-quantitative analysis of the solutions of the dynamical equations and their stability. DSA has been widely used in gravitational theories \cite{wainwright_ellis_1997,Carloni:2013hna,Bahamonde:2021gfp,Dutta:2017fjw,Odintsov:2018uaw}, including the $f(\mathcal{Q})$ theory \cite{Narawade:2022jeg,Lu:2019hra,Khyllep:2021pcu}. The $1+3$ approach provides an ideal framework to employ the DSA, as it makes available convenient variables for the description of the phase space of the dynamical system. As we will see, in the context of the Bianchi type-I metric, the application of the $1+3$ approach leads to a remarkable simplification of the involved equations. For example, we will be able to deal with complex matter sources as well as non-trivial forms of the function $f(\mathcal{Q})$.

The paper is organized as follows. Sec. \ref{sec:geometric_framework} is devoted to some geometrical preliminaries. Sec. \ref{sec:1_3_description} introduces the framework of the $1+3$ approach, defining all the necessary kinematic quantities and providing the time and space decomposition with respect to the given congruence of all the relevant geometrical quantities. A review of $f(\mathcal{Q})$ gravity and its cosmological equations is given in Sec. \ref{sec:f_Q_theory}, whereas Bianchi type-I universes are discussed in Sec. \ref{sec:bianchi_model}. In Sec. \ref{sec:dynamical_systems}, DSA is applied to investigate four different cosmological scenarios. Eventually, the obtained results are discussed in Sec. \ref{sec:conclusions}. 

Throughout the paper natural units ($c=8\pi G=1$) and metric signature ($-,+,+,+$) are used.


\section{Geometrical preliminaries}\label{sec:geometric_framework}
We consider a spacetime endowed with a metric tensor $g_{ij}$ and a torsion free affine connection $\Gamma_{ij}{}^{k}$. The latter can be decomposed as:
\begin{equation}\label{eq:def_connection}
\Gamma_{ij}{}^{k} = \tilde{\Gamma}_{ij}{}^{k} + N_{ij}{}^{k} ,
\end{equation}
where $\tilde{\Gamma}_{ij}{}^{k}$ is the Levi-Civita connection induced by the metric tensor $g_{ij}$,
\begin{equation}\label{eq:def_levicivita}
\tilde{\Gamma}_{ij}{}^{k} = \frac{1}{2} g^{kh} \left( \partial_{i}g_{jh} + \partial_{j}g_{ih} - \partial_{h}g_{ij} \right),
\end{equation}
and $N_{ij}{}^{k}$ is the disformation tensor,
\begin{equation}\label{eq:def_disformation}
N_{ij}{}^{k} = \frac{1}{2} \left( Q^{k}{}_{ij} - Q_{i}{}^{k}{}_{j} - Q_{j}{}^{k}{}_{i} \right),
\end{equation}
defined in terms of the nonmetricity tensor,
\begin{equation}\label{eq:def_nonmetricity}
Q_{kij} = \nabla_{k}g_{ij},
\end{equation}
being $\nabla$ the covariant derivative associated with the full connection \eref{eq:def_connection}. Throughout the paper, we will denote by a tilde all quantities related to the Levi-Civita connection. For instance, we will indicate by $\tilde{\nabla}$ the covariant derivative associated with the Levi-Civita connection \eref{eq:def_levicivita}.

The curvature tensor of the full connection \eref{eq:def_connection} is defined as
\begin{equation}\label{eq:def_riemann_1}
    \eqalign{
     R^{h}{}_{kij} &= \partial_{i}\Gamma_{jk}{}^{h} - \partial_{j}\Gamma_{ik}{}^{h} + \Gamma_{ip}{}^{h}\Gamma_{jk}{}^{p} - \Gamma_{jp}{}^{h}\Gamma_{ik}{}^{p} =\\
    &= \tilde{R}^{h}{}_{kij} + \tilde{\nabla}_{i}N_{jk}{}^{h} - \tilde{\nabla}_{j}N_{ik}{}^{h} + N_{ip}{}^{h}N_{jk}{}^{p} - N_{jp}{}^{h}N_{ik}{}^{p}
    }
\end{equation}
according to the Ricci identity,
\begin{equation}\label{eq:def_riemann_0}
    R^{h}{}_{kij}w^{k} = \left(\nabla_{i}\nabla_{j}-\nabla_{j}\nabla_{i}   \right)w^{h},
\end{equation}
where $w^{h}$ is a generic vector field. In the presence of nonmetricity, we recall that the Riemann tensor  \eref{eq:def_riemann_1} satisfies the properties:
\begin{itemize}\label{eq:riemann_properties}
    \item Antisymmetry in the last two indices,
    \begin{equation}
       R^{h}{}_{kij} = -R^{h}{}_{kji};
    \end{equation}
    \item First Bianchi identity,
    \begin{equation}
       R^{h}{}_{[kij]}=0;
    \end{equation}
    \item Second Bianchi identity,
    \begin{equation}
        \nabla_{[a|}R^{h}{}_{k|ij]}=0;
    \end{equation}
\end{itemize}
By contracting first and third index of the Riemann tensor, we obtain the Ricci tensor,
\begin{equation}\label{eq:def_ricci_tensor}
    R_{kj} = R^{h}{}_{khj}.
\end{equation}
The contraction between second and third index gives rise to,
\begin{equation}\label{eq:def_second_contraction_riemann}
    \bar{R}_{hj} = R_{h}{}^{i}{}_{ij}.
\end{equation}
Also, the contraction of first and second index yields the homothetic curvature,
\begin{equation}\label{eq:def_homothetic_tensor}
    \hat{R}_{ij} = R^{h}{}_{hij} = - \tilde{\nabla}_{[i}Q_{j]h}{}^{h} = - \partial_{[i}Q_{j]h}{}^{h}.
\end{equation}
Eqs. \eref{eq:def_ricci_tensor} and \eref{eq:def_second_contraction_riemann} contracted with the metric give the Ricci scalar,
\begin{equation}\label{eq:def_ricci_scalar}
    R= g^{ij}R_{ij} = - g^{ij}\bar{R}_{ij}.
\end{equation}


\section{\texorpdfstring{$1+3$ framework}{}}\label{sec:1_3_description}

In this section, we apply the $1+3$ formalism to the spacetime described in Sec. \ref{sec:geometric_framework}. The approach is based on the introduction of a congruence of time-like curves, or world lines, representing preferred observers. The aim is to analyze how nonmetricity affects them. 

\subsection{\texorpdfstring{$4-\hbox{velocity}$}{}}
Given the congruence $x^{i}=x^{i}(\lambda)$, expressed in terms of an affine parameter $\lambda$, we define the $4$-velocity as the time-like vector:
\begin{equation}
    u^{i} = \frac{d x^{i}}{d\lambda}, \qquad u_{i}=g_{ij}u^{j}.
\end{equation}
However, due to nonmetricity, in general the proper time $\tau$ is not an affine parameter. In this regard, it is easily seen that the condition
\begin{equation}\label{eq:condition_about_u}
    Q_{kij}u^{k}u^{i}u^{j} = 0, 
\end{equation} 
together with the requirement for a curve to be autoparallel, ensures that proper time is actually an affine parameter \cite{Iosifidis:2018diy}. Assuming systematically the condition \eref{eq:condition_about_u}, we can arrange things in order to parameterize the curves using the proper time and define the $4$-velocity as,
\begin{equation}\label{eq:def_4_velocity}
    u^{i} = \frac{d x^{i}}{d\tau}, \qquad u_{i}u^{i} = -1.
\end{equation}
As we will deal exclusively with cosmological models of Bianchi type-I, we will assume in Sec. \ref{sec:bianchi_model} that the condition \eref{eq:condition_about_u} is satisfied. Indeed, in \ref{appendix:bianchi_coordinates} we will show that, because of the gauge choice, such a condition is not restrictive for our purposes.

Once the $4$-velocity has been defined, we may introduce the projection operator along $u^{i}$, defined by means of the tensor, 
\begin{equation}\label{eq:4_velocity_projection}
    U^{i}{}_{j}=-u^{i}u_{j},
\end{equation}
satisfying the properties,
\begin{equation}
    U^{i}{}_{j}u^{j} = u^{i}, \qquad U^{i}{}_{k}U^{k}{}_{j} = U^{i}{}_{j}, \qquad U^{i}{}_{i} = 1.
\end{equation}

\subsubsection{Orthogonal projection}
The choice of a preferred time direction allows us to single out a three-dimensional subspace of the tangent bundle at any point, orthogonal to the $4$-velocity $u^i$. The restriction of the metric to this spatial subspace is the so-called transverse metric,
\begin{equation}\label{eq:def_3_metric}
    h_{ij} = g_{ij} + u_{i}u_{j}.
\end{equation}
Associated with the transverse metric \eref{eq:def_3_metric} there is the spatial projection operator,
\begin{equation}\label{eq:def_orthogonal_projection}
    h^{i}{}_{j}=\delta^{i}_{j} + u^{i}u_{j},
\end{equation}
satisfying the properties
\begin{equation}
    h^{i}{}_{k}h^{k}{}_{j}=h^{i}{}_{j}, \qquad h^{i}{}_{i}=3, \qquad h^{i}{}_{j}u^{j}=0.
\end{equation}
In the following discussion, we will also use the projected symmetric trace free (PSTF) part of a tensor. In particular, for any $1$-form $V_{i}$ and covariant $2$-tensor $T_{ij}$, the PSTF is expressed as,
\begin{equation}\label{eq:def_PSTF}
    V_{\langle  i\rangle  } = h_{i}{}^{j}V_{j}, \qquad T_{\langle ij\rangle  } = \left[h_{(i}{}^{m}h_{j)}{}^{n} - \frac{1}{3}h_{ij}h^{mn}\right]T_{mn}.
\end{equation}

\subsection{Time and spatial derivative}\label{sec:time_derivative}
The time derivative of a generic tensor $T^{i\cdot\cdot\cdot}{}_{j\cdot\cdot\cdot}$ is defined as,
\begin{equation}\label{eq:def_time_derivative}
\eqalign{
    \fl \dot{T}^{i\cdot\cdot\cdot}{}_{j\cdot\cdot\cdot} &= u^{h}\nabla_{h}T^{i\cdot\cdot\cdot}{}_{j\cdot\cdot\cdot}=\\ 
    \fl &= u^{h}\tilde{\nabla}_{h}T^{i\cdot\cdot\cdot}{}_{j\cdot\cdot\cdot} + u^{h}N_{hk}{}^{i}T^{k\cdot\cdot\cdot}{}_{j\cdot\cdot\cdot}+\cdot\cdot\cdot-u^{h}N_{hj}{}^{k}T^{i\cdot\cdot\cdot}{}_{k\cdot\cdot\cdot}-\cdot\cdot\cdot=\\
    \fl &= \mathring{{T}}^{i\cdot\cdot\cdot}{}_{j\cdot\cdot\cdot} + u^{h}N_{hk}{}^{i}T^{k\cdot\cdot\cdot}{}_{j\cdot\cdot\cdot}+\cdot\cdot\cdot-u^{h}N_{hj}{}^{k}T^{i\cdot\cdot\cdot}{}_{k\cdot\cdot\cdot}-\cdot\cdot\cdot,
    }
\end{equation}
where
\begin{equation}\label{eq:def_levi_civita_time_derivative}
     \mathring{T}^{i\cdot\cdot\cdot}{}_{j\cdot\cdot\cdot} = u^{h}\tilde{\nabla}_{h}T^{i\cdot\cdot\cdot}{}_{j\cdot\cdot\cdot}
\end{equation}
is the time derivative with respect to the Levi-Civita connection. 

The spatial derivative is the spatial projection of the covariant derivative,
\begin{eqnarray}
   \fl D_{k}T^{i\cdot\cdot\cdot}{}_{j\cdot\cdot\cdot} &=& h_{k}{}^{p} h^{i}{}_{m}\cdot\cdot\cdot h_{j}{}^{n}\cdot\cdot\cdot \nabla_{p}T^{m\cdot\cdot\cdot}{}_{n\cdot\cdot\cdot} =\nonumber\\
   \fl &=& h_{k}{}^{p} h^{i}{}_{m} \cdot\cdot\cdot h_{j}{}^{n} \cdot\cdot\cdot \left(\tilde{\nabla}_{p}T^{m\cdot\cdot\cdot}{}_{n\cdot\cdot\cdot} + N_{pq}{}^{m}T^{q\cdot\cdot\cdot}{}_{n\cdot\cdot\cdot} + \cdot\cdot\cdot +\nonumber\right.\\
   \fl &&\left.- N_{pn}{}^{q}T^{m\cdot\cdot\cdot}{}_{q\cdot\cdot\cdot}-\cdot\cdot\cdot\right)=\nonumber\\
   \fl &=& \tilde{D}_{k}T^{i\cdot\cdot\cdot}{}_{j\cdot\cdot\cdot} + h_{k}{}^{p} h^{i}{}_{m} \cdot\cdot\cdot h_{j}{}^{n} \cdot\cdot\cdot N_{pq}{}^{m}T^{q\cdot\cdot\cdot}{}_{n\cdot\cdot\cdot} +\nonumber\\
   \fl &&+ \cdot\cdot\cdot - h_{k}{}^{p} h^{i}{}_{m} \cdot\cdot\cdot h_{j}{}^{n} \cdot\cdot\cdot N_{pn}{}^{q}T^{m\cdot\cdot\cdot}{}_{q\cdot\cdot\cdot}-\cdot\cdot\cdot,
   \label{eq:def_spatial_derivative}
\end{eqnarray}
with
\begin{equation}
    \tilde{D}_{k}T^{i\cdot\cdot\cdot}{}_{j\cdot\cdot\cdot} = h_{k}{}^{p} h^{i}{}_{m} \cdot\cdot\cdot h_{j}{}^{n} \cdot\cdot\cdot \tilde{\nabla}_{p}T^{m\cdot\cdot\cdot}{}_{n\cdot\cdot\cdot}
\end{equation}
the spatial derivative with respect to the Levi-Civita connection. It is worth noticing that the spatial derivative of the metric $g_{ij}$ is equal to the spatial derivative of $h_{ij}$:
\begin{equation}\label{eq:spatial_derivative_metric}
    \fl D_{k}g_{ij} = h_{k}{}^{p} h_{i}{}^{m} h_{j}{}^{n} \nabla_{p}g_{mn} = h_{k}{}^{p} h_{i}{}^{m} h_{j}{}^{n} \nabla_{p}\left(h_{mn} - u_{m}u_{n}\right) = D_{k}h_{ij}.
\end{equation}

\subsubsection{\texorpdfstring{$4-\hbox{acceleration}$}{}}\label{sec:4_acceleration}
Because of nonmetricity, scalar product and covariant derivative do not commute in general. For this reason, we use the convention that the contravariant, or covariant, counterparts of objects related to the covariant derivative are obtained raising, or lowering, the indices by the metric. This convention will be used throughout the paper. Accordingly, we define the $4$-acceleration as
\begin{equation}\label{eq:def_cov_acceleration}
    \dot{u}_{i} = u^{h}\nabla_{h}u_{i} = \mathring{u}_{i} - N_{hi}{}^{k}u_{k}u^{h} = \mathring{{u}}_{i} + \frac{1}{2}Q_{ihk}u^{h}u^{k}.
\end{equation}
where $\mathring{{u}}_{i}:=u^{h}\tilde{\nabla}_{h}u_{i}$ is the $4$-acceleration with respect to the Levi-Civita connection. After that, the contravariant counterpart of \eref{eq:def_cov_acceleration} is obtained as
\begin{equation}\label{eq:def_vec_acceleration}
    \dot{u}^{i} := g^{ij} \dot{u}_{j} = g^{ij}u^{h}\nabla_{h}u_{j} = \mathring{{u}}^{i} + \frac{1}{2}g^{ij}Q_{jhk}u^{h}u^{k}.
\end{equation}
If Eq. \eref{eq:condition_about_u} holds, $u_{i}$ and $\dot{u}_{i}$ are orthogonal to each other,
\begin{equation}
   \dot{u}_{i}u^{i} = \dot{u}^{i}u_{i} = 0.
\end{equation}

\subsubsection{Extrinsic curvature}
The extrinsic curvature is defined as the spatial derivative of $4$-velocity,
\begin{equation}\label{eq:def_extrinsic_curvature}
    K_{ij} = D_{i}u_{j} = h_{i}{}^{m}h_{j}{}^{n}\nabla_{m}u_{n} = \tilde{K}_{ij} - h_{i}{}^{m}h_{j}{}^{n}N_{mn}{}^{h}u_{h},
\end{equation}
being
\begin{equation}\label{eq:def_levi_civita_extrinsic_curvature}
    \tilde{K}_{ij} = \tilde{D}_{i}u_{j}
\end{equation}
the extrinsic curvature induced by the Levi-Civita connection. Raising the second index, we obtain 
\begin{equation}
    K_{i}{}^{j}= g^{jp}K_{ip} = \tilde{K}_{i}{}^{j} - g^{jp}h_{i}{}^{m}h_{p}{}^{n}N_{mn}{}^{k}u_{k} . 
\end{equation}

\subsection{Kinematic quantities}
The covariant derivative of the $4$-velocity can be decomposed in its temporal and spatial projections,
\begin{equation}\label{eq:covariant_derivative_4_velocity}
\eqalign{
    \nabla_{i}u_{j} &= -u_{i}\dot{u}_{j} + D_{i}u_{j} - u_{j}h_{i}{}^{k}u^{l}\nabla_{k}u_{l}=\\ 
    &= -u_{i}\dot{u}_{j} + \frac{1}{3}h_{ij}\Theta + \sigma_{ij} + \omega_{ij} - \frac{1}{2}u_{j}h_{i}{}^{k}Q_{kmn}u^{m}u^{n},
    }  
\end{equation}
with 
\begin{equation}
    D_{i}u_{j} := \frac{1}{3}h_{ij}\Theta + \sigma_{ij} + \omega_{ij},
\end{equation}
and where:
\begin{itemize}
\item $\Theta$ is related to the rate of volume expansion,
\begin{equation}\label{eq:def_Theta}
    \fl \Theta = g^{ij} D_{i}u_{j} = g^{ij}h_{i}{}^{p}h_{j}{}^{q}\nabla_{p}u_{q} = h^{ij}\tilde{D}_{i}u_{j} - h^{ij}N_{ij}{}^{k}u_{k} = \tilde{\Theta} - h^{ij}N_{ij}{}^{k}u_{k},
\end{equation}
with
\begin{equation}\label{eq:def_Theta_levi_civita}
    \tilde{\Theta} = \tilde{D}_{i}u^{i};
\end{equation}
\item $\sigma_{ij}$ is the trace-free symmetric tensor called ``shear tensor'', describing the volume preserving distortion of the fluid flow,
\begin{equation}\label{eq:def_sigma}
\eqalign{
    \fl \sigma_{ij} = D_{ \langle i}u_{j \rangle} &= \left[ h_{(i}{}^{m}h_{j)}{}^{n} - \frac{1}{3}h_{ij}h^{mn}\right] \left( \tilde{D}_{m}u_{n} - h_{m}{}^{p}h_{n}{}^{q}N_{pq}{}^{k}u_{k} \right) =\\
    \fl &= \tilde{\sigma}_{ij} - 
    N_{\langle ij \rangle}{}^{k}u_{k},}
\end{equation}
\begin{equation}
    \fl \sigma_{ij}u^{j}=0, \qquad \sigma_{i}{}^{i}=0,
\end{equation}
with
\begin{equation}\label{eq:def_sigma_levi_civita}
    \fl \tilde{\sigma}_{ij} = \tilde{D}_{\langle i}u_{j\rangle  }, \qquad \tilde{\sigma}_{ij}u^{j}=0, \qquad  \tilde{\sigma}_{i}{}^{i}= 0;
\end{equation}
\item $\omega_{ij}$ is the skew-symmetric  tensor called ``vorticity tensor'' describing rotation of the fluid flow,
\begin{equation}\label{eq:def_omega}
    \fl \omega_{ij} = D_{[i}u_{j]} = \tilde{D}_{[i}u_{j]} - h_{[i}{}^{m}h_{j]}{}^{n}N_{mn}{}^{k}u_{k} = \tilde{D}_{[i}u_{j]} = \tilde{\omega}_{ij},
\end{equation}
\begin{equation}\label{eq:def_omega_levi_civita}
    \fl \tilde{\omega}_{ij} = \tilde{D}_{[i}u_{j]}, \qquad \omega_{ij}u^{j}=\tilde{\omega}_{ij}u^{j}=0.
\end{equation}
\end{itemize}
It is useful to introduce the magnitudes of shear and vorticity tensors:
\begin{equation}
    \sigma^{2} = \frac{1}{2}\sigma_{ij}\sigma^{ij}, \qquad \omega^{2} = \frac{1}{2}\omega_{ij}\omega^{ij},
\end{equation}
\begin{equation}
    \tilde{\sigma}^{2} = \frac{1}{2}\tilde{\sigma}_{ij}\tilde{\sigma}^{ij}, \qquad \tilde{\omega}^{2} = \frac{1}{2}\tilde{\omega}_{ij}\tilde{\omega}^{ij}.
\end{equation}
Substituting Eqs. \eref{eq:def_Theta}, \eref{eq:def_sigma} and \eref{eq:def_omega} in Eq. \eref{eq:covariant_derivative_4_velocity} and considering Eq. \eref{eq:condition_about_u}, we get the expression,
\begin{equation}\label{eq:covariant_derivative_4_velocity_levi_civita}
\eqalign{
    \fl \nabla_{i}u_{j} &= \tilde{\nabla}_{i}u_{j} - \frac{1}{2}u_{i}h_{j}{}^{k}Q_{kmn}u^{m}u^{n} - h_{i}{}^{m}h_{j}{}^{n}N_{mn}{}^{p}u_{p} - \frac{1}{2}u_{j}h_{i}{}^{k}Q_{kmn}u^{m}u^{n} =\\
	\fl &= -u_{i}\mathring{{u}}_{j} +\frac{1}{3}\tilde{\Theta} h_{ij} + \tilde{\sigma}_{ij} + \tilde{\omega}_{ij} - u_{(i}h_{j)}{}^{k}Q_{kmn}u^{m}u^{n} - h_{i}{}^{m}h_{j}{}^{n}N_{mn}{}^{p}u_{p}.
    }
\end{equation}

\subsection{Gauss Relation}
Given a spatial vector field $v^{p}$, we define the spatial Riemann tensor through the relation,
\begin{equation}\label{eq:def_3_riemann}
\, ^{3}R^{p}{}_{qij}v^{q} :=   \left(D_{i}D_{j}-D_{j}D_{i}\right)v^{p} -2 \omega_{ij}u^{r}h_{s}{}^{p}\nabla_{r}v^{s}.
\end{equation}
This relation can be recast as the so-called ``Gauss relation'',
\begin{equation}\label{eq:gauss_relation}
    \fl \, ^{3}R^{p}{}_{qij} = h_{i}{}^{m}h_{j}{}^{n}h^{p}{}_{s}h_{q}{}^{r}R^{s}{}_{rmn} + K_{j}{}^{p}K_{iq} - K_{i}{}^{p}K_{jq} +2 h_{[i}{}^{m}K_{j]q}h^{pn}Q_{mns}u^{s}.
\end{equation}
Since Riemann tensor is not antisymmetric in the first two indices, we can obtain two ``contracted Gauss relations''. The first contracting the first and third index of the Gauss relation,
\begin{equation}\label{eq:contracted_gauss_relation_1}
    \fl \, ^{3}R_{qj} = \, ^{3}R^{i}{}_{qij} = h_{s}{}^{m}h_{j}{}^{n}h_{q}{}^{r}R^{s}{}_{rmn} + K_{j}{}^{i}K_{iq} - K_{i}{}^{i}K_{jq} + 2 h_{[i}{}^{m}K_{j]q}h^{in}Q_{mns}u^{s},
\end{equation}
and a second one contracting the second and third indices,
\begin{equation}\label{eq:contracted_gauss_relation_2}
    \fl \, ^{3}\bar{R}_{qj} = \, ^{3}R_{q}{}^{i}{}_{ij} = h_{r}{}^{m}h_{j}{}^{n}h_{q}{}^{s}R_{s}{}^{r}{}_{mn} + K_{jq}K_{i}{}^{i} - K_{iq}K_{j}{}^{i} + 2 h_{[i}{}^{m}K_{j]}{}^{i}h_{q}{}^{n}Q_{mns}u^{s}.
\end{equation}
The trace of both Eqs. \eref{eq:contracted_gauss_relation_1} and Eq. \eref{eq:contracted_gauss_relation_2} leads to the ``scalar Gauss relation'',
\begin{equation}\label{eq:scalar_gauss_relation}
\eqalign{
    \fl \, ^{3}R &= g^{qj}\, ^{3}R_{qj} = - g^{qj}\, ^{3}\bar{R}_{qj} =\\
    \fl &= h_{s}{}^{m}h^{rn}R^{s}{}_{rmn} + K^{ji}K_{ij} - K_{i}{}^{i}K_{j}{}^{j} + 2 h_{[i}{}^{m}K_{j]}{}^{j}h^{in}Q_{mns}u^{s},
    }  
\end{equation}
which generalizes the ``Theorema Egregium'' in the presence of nonmetricity.

It is also useful to rewrite Eq. \eref{eq:gauss_relation} in the form
\begin{equation}\label{eq:gauss_relation_levi_civita}
\eqalign{
    \fl \, ^{3}R^{p}{}_{qij} =& \, ^{3}\tilde{R}^{p}{}_{qij} + 2 \tilde{D}_{[i}N_{j]q}{}^{p} + 2 \tilde{K}_{[i}{}^{p}h_{j]}{}^{m}h_{q}{}^{n}N_{mn}{}^{k}u_{k} +\\
    \fl &+ 2 \tilde{K}_{[i|q}h_{|j]}{}^{m}h^{p}{}_{n}N_{mk}{}^{n}u^{k} + 2 h_{[i}{}^{r}h_{j]}{}^{m}h_{s}{}^{p}h_{q}{}^{n}h_{l}{}^{k}N_{rk}{}^{s}N_{mn}{}^{l},
    }    
\end{equation}
in which the contributions due to Levi-Civita and nonmetricity terms are made evident.

\subsection{Energy-momentum tensor}
The energy-momentum tensor of the matter fluid can be decomposed in its irreducible parts as,
\begin{equation}\label{eq:energy_momentum_tensor}
    \Psi_{ij} = \rho u_{i}u_{j} + q_{i}u_{j} + u_{i}q_{j} + p h_{ij} + \pi_{ij},
\end{equation}
where 
\begin{equation}
    \rho = \Psi_{ij}u^{i}u^{j}
\end{equation}
is the relativistic energy density,
\begin{equation}
    q_{i} = - h_{i}{}^{k}\Psi_{kj}u^{j}
\end{equation}
the relativistic energy flux,
\begin{equation}
    p = \frac{1}{3}h^{ij}\Psi_{ij}
\end{equation}
the isotropic pressure, and
\begin{equation}
    \pi_{ij} = \Psi_{\langle ij \rangle}
\end{equation}
the trace-free  anisotropic pressure.
The trace of tensor \eref{eq:energy_momentum_tensor} is equal to,
\begin{equation}
    \Psi = \Psi_{i}{}^{i} = -\rho + 3p.
\end{equation}

\subsection{Nonmetricity decomposition}\label{sec:nonmetricity_decomposition}
Similarly to what we have done for the energy-momentum tensor \eref{eq:energy_momentum_tensor}, we can decompose the nonmetricity tensor using $u_i$ and $h_{ij}$ as:
\begin{equation}\label{eq:non_metricity_decomposition}
\eqalign{
    \fl Q_{kij} =& -Q_{0}u_{k}u_{i}u_{j} - \frac{1}{3} Q_{1} u_{k}h_{ij} - \frac{2}{3}Q_{2}u_{(i}h_{j)k} + Q^{(0)}{}_{k}u_{i}u_{j} + 2Q^{(1)}{}_{(i}u_{j)}u_{k} + \\
    \fl &+ \frac{1}{3}Q^{(2)}{}_{k}h_{ij} + \frac{2}{3}Q^{(3)}{}_{(i}h_{j)k} - Q^{(0)}{}_{ij}u_{k} - 2Q^{(1)}{}_{k(i}u_{j)} + \, ^{3}Q_{kij},
    }
\end{equation}
where 
\begin{equation}
    Q_{0}= Q_{kij}u^{k}u^{i}u^{j}, \quad Q_{1} = Q_{kij}u^{k}h^{ij}, \quad Q_{2} = Q_{kij}h^{ki}u^{j} 
\end{equation}
are scalar quantities,
\begin{equation}
    Q^{(0)}{}_{k} = Q_{pij} h^{p}{}_{k}u^{i}u^{j}, \qquad Q^{(1)}{}_{k} = Q_{pij}u^{p}u^{j}h^{i}{}_{k},
\end{equation}
\begin{equation}
    Q^{(2)}{}_{k} = Q_{pij}h^{p}{}_{k}h^{ij}, \qquad Q^{(3)}{}_{k} = Q_{pij} h^{i}{}_{k} h^{pj}
\end{equation}
are covectors,
\begin{equation}
    Q^{(0)}{}_{ij} = Q^{(0)}{}_{ji} = \left[h_{(i}{}^{p}h_{j)}{}^{q} - \frac{1}{3}h_{ij}h^{pq}\right]Q_{kpq} u^{k},
\end{equation}
\begin{equation} 
    Q^{(1)}{}_{ij}=\left(h_{i}{}^{p}h_{j}{}^{q} - \frac{1}{3}h_{ij}h^{pq}\right)Q_{pkq} u^{k}
\end{equation}
are trace-free tensors and
\begin{equation}
    \fl \, ^{3}Q_{kij} = h_{k}{}^{p}h_{i}{}^{q}h_{j}{}^{r}Q_{pqr} - \frac{1}{3}h_{ij}h^{qr}h_{k}{}^{p}Q_{pqr} - \frac{1}{3}h_{ki}h^{pq}h_{j}{}^{r}Q_{pqr} - \frac{1}{3}h_{kj}h^{pr}h_{i}{}^{q}Q_{pqr}
\end{equation}
is a fully spatial tensor, whose traces are given by
\begin{equation}
    \, ^{3}Q_{ji}{}^{j} = -\frac{1}{3}Q^{(2)}{}_{i} -\frac{1}{3}Q^{(3)}_{i}  \quad \hbox{and} \quad  \, ^{3}Q_{ij}{}^{j} = - \frac{2}{3}Q^{(3)}{}_{i}.
\end{equation}
However, unlike the energy-momentum tensor, Eq. \eref{eq:non_metricity_decomposition} is not an irreducible decomposition. Moreover, because of Eq. \eref{eq:condition_about_u}, $Q_{0}=0$.

We can now rewrite Eqs. \eref{eq:def_cov_acceleration}, \eref{eq:def_Theta}, and \eref{eq:def_sigma} in terms of different contributions of nonmetricity:
\begin{equation}\label{eq:4_velocity_non_metricity_decomposition}
    \dot{u}_{i} = \mathring{u}_{i} + \frac{1}{2} Q^{(0)}{}_{i},
\end{equation}
\begin{equation}\label{eq:4_velocity_non_metricity_decomposition_2}
    \Theta = \tilde{\Theta} - \frac{1}{2}Q_{1} + Q_{2},
\end{equation}
\begin{equation}\label{eq:sigma_nonmetricity_decomposition}
    \sigma_{ij} = \tilde{\sigma}_{ij} - \frac{1}{2}Q^{(0)}{}_{ij} + Q^{(1)}{}_{( ij )}.
\end{equation}
Eqs. \eref{eq:4_velocity_non_metricity_decomposition}, \eref{eq:4_velocity_non_metricity_decomposition_2} and \eref{eq:sigma_nonmetricity_decomposition} show how nonmetricity affects the kinematic quantities associated with the given congruence.


\section{\texorpdfstring{$f(\mathcal{Q})$}{} theory}\label{sec:f_Q_theory}
$f(\mathcal{Q})$ gravity is a generalization of Symmetric Teleparallel Gravity, where the gravitational Lagrangian $f(\mathcal{Q})$ is a given function of the nonmetricity scalar. The latter is defined as,
\begin{equation}\label{eq:def_non_metricity_scalar}
\eqalign{
    \mathcal{Q} &= N_{hp}{}^{h}N_{k}{}^{kp} - N_{kp}{}^{h}N_{h}{}^{kp} = -Q_{hij}P^{hij} =\\
    &= \frac{1}{4}Q_{hij}Q^{hij} - \frac{1}{2}Q_{hij}Q^{ijh} - \frac{1}{4} q_{h}q^{h} + \frac{1}{2}q_{h}Q^{h},
    }
\end{equation}
where
\begin{equation}
    P^{h}{}_{ij} = - \frac{1}{4} Q^{h}{}_{ij} + \frac{1}{2} Q_{(ij)}{}^{h} + \frac{1}{4} q^{h} g_{ij} - \frac{1}{4} Q^{h} g_{ij} - \frac{1}{4} \delta^{h}_{(i} q_{j)}
\end{equation}
is the conjugate tensor of $Q_{hij}$, and 
\begin{equation}
    q_{h} = Q_{hi}{}^{i} \qquad Q_{h} = Q_{ih}{}^{i}
\end{equation}
are its two independent traces. Writing the Ricci scalar in the form,
\begin{equation}\label{eq:ricci_scalar_Q}
    R = \tilde{R} + \tilde{\nabla}_{h}N_{k}{}^{kh}- \tilde{\nabla}_{k}N_{h}{}^{kh} + \mathcal{Q},
\end{equation}
allows us to  highlight how the Lagrangians of STG and GR differ by a total divergence (and a sign).

In a metric-affine framework, the field equations of $f(\mathcal{Q})$ gravity are derived from the action
\begin{equation}\label{eq:f(Q)_action}
    \fl A = \int \hbox{d}^4x \left[-\frac{1}{2} \sqrt{-g} f(\mathcal{Q}) + \lambda_{a}{}^{bij}R^{a}{}_{bij} + \lambda_{a}{}^{ij}T_{ij}{}^{a} + \sqrt{-g}\mathcal{L}_{m}\right],
\end{equation}
where $\mathcal{L}_{m}$ is the matter Lagrangian, $\lambda_{a}{}^{bij}$ and $\lambda_{a}{}^{ij}$ are Lagrange multipliers introduced to impose the vanishing of curvature and torsion. Performing variations, we get
\begin{equation}
    R^{h}{}_{kij} = 0, \qquad  T_{ij}{}^{h} = 0,
\end{equation}
\begin{equation} \label{eq:metric_equation}
    \fl \frac{2}{\sqrt{-g}}\nabla_{h} \left( \sqrt{-g} f' P^{h}{}_{ij} \right) + \frac{1}{2}g_{ij}f(\mathcal{Q}) + f' \left( P_{ihk}Q_{j}{}^{hk} - 2 Q^{hk}{}_{i}P_{hkj} \right) = \Psi_{ij}, 
\end{equation}
and
\begin{equation}\label{eq:connection_equation}
    \nabla_{i}\nabla_{j}\left(\sqrt{-g} f' P^{ij}{}_{h}\right) +\nabla_{i}\nabla_{j} \Phi^{ij}{}_{h}=0,
\end{equation}
with 
\begin{equation}
    \Psi_{ij}=  -\frac{2}{\sqrt{-g}}\frac{\delta \left(\sqrt{-g} \mathcal{L}_{m}\right)}{\delta g^{ij}} \:\: \hbox{and} \:\:
    \Phi^{ij}{}_{h} = - \frac{1}{2}\frac{\delta \left( \sqrt{-g}\mathcal{L}_{m}\right)}{\delta {\Gamma_{ij}{}^{h}}}.
\end{equation}
From Eq. \eref{eq:connection_equation} and the Levi-Civita divergence of Eq. \eref{eq:metric_equation}, we derive the energy-momentum conservation law,
\begin{equation}\label{eq:energy-momentum_conservation}
\tilde{\nabla}_{i}\Psi^{i}{}_{h} + \frac{2}{\sqrt{-g}}\nabla_{i}\nabla_{j} \Phi^{ij}{}_{h} = 0.
\end{equation}
Since we consider matter independent of nonmetricity, $\Phi^{ij}{}_{h}$ is identically zero.

In view of condition $R^{a}{}_{bij} = 0$, Eq. \eref{eq:metric_equation} can be reformulated more suitably for the $1+3$ formalism as (see \ref{appendix:field_equations_derivation}):
\begin{equation}\label{eq:final_field_equation}
    \fl \tilde{R}_{ij} = \frac{1}{f'} \left( \Psi_{ij} - \frac{1}{2}g_{ij}\Psi \right) + \frac{1}{2}g_{ij}\left(\frac{f}{f'} - \mathcal{Q}\right) - 2 \frac{f''}{f'}\left( P^{h}{}_{ij}-\frac{1}{2}g_{ij}P^{hk}{}_{k} \right)\partial_{h}\mathcal{Q}.
\end{equation}
By replacing $f(\mathcal{Q}) = \mathcal{Q}$ into Eq. \eref{eq:final_field_equation}, we recover the field equations of General Relativity,
\begin{equation}
    \tilde{R}_{ij} = \Psi_{ij} - \frac{1}{2}g_{ij}\Psi.
\end{equation}

\subsection{Cosmological equations}
At this point, making use of the following relations for $\tilde{R}_{ij}$,
\begin{equation}\label{eq:ricci_u_u_levi_civita}
    \fl \tilde{R}_{ij}u^{i}u^{j} = u^{j}\tilde{\nabla}_{h}\tilde{\nabla}_{j}u^{h} - \left( \tilde{\nabla}_{h}u^{h}\right)^{\cdot} =  -\mathring{\tilde{\Theta}}-\frac{1}{3}\tilde{\Theta}^{2}-2\left(\tilde{\sigma}^{2}-\tilde{\omega}^{2}\right) + \tilde{D}_{h}\mathring{u}^{h} + \mathring{u}^{h}\mathring{u}_{h},
\end{equation}
and
\begin{equation}\label{eq:gauss_ricci_levi_civita}
    \fl \, ^{3}\tilde{R}_{ij} = h_{j}{}^{q}h_{i}{}^{p} \tilde{R}_{pq} - \tilde{K}_{p}{}^{p}\tilde{K}_{ji} - h_{j}{}^{q}h_{i}{}^{p}u^{m}\tilde{\nabla}_{m}\tilde{K}_{qp} + \tilde{D}_{j}\mathring{{u}}_{i} + \mathring{{u}}_{i}\mathring{{u}}_{j},
\end{equation}
we can derive the $1+3$ cosmological equations for a generic $f(\mathcal{Q})$ theory, namely:
\begin{itemize}
    \item Raychaudhuri equation, obtained from Eq. \eref{eq:ricci_u_u_levi_civita},
        \begin{equation}\label{eq:raychaudhuri_equation}
        \eqalign{
        \fl \mathring{\tilde{\Theta}} + \frac{1}{3}\tilde{\Theta}^{2} +& 2 \left(\tilde{\sigma}^{2} - \tilde{\omega}^{2}\right) -\tilde{D}_{i}\mathring{{u}}^{i} - \mathring{{u}}^{i}\mathring{{u}}_{i} + \frac{1}{2f'}\left( \rho + 3p \right) +\\
        \fl &- \frac{1}{2}\left(\frac{f}{f'} - \mathcal{Q}\right) - 2 \frac{f''}{f'}\left( P^{h}{}_{ij}u^{i}u^{j} + \frac{1}{2}P^{hk}{}_{k} \right)\partial_{h}\mathcal{Q} = 0; \
        }  
        \end{equation}
    \item Spatial equations,  derived from Eqs. \eref{eq:gauss_ricci_levi_civita} and \eref{eq:raychaudhuri_equation}, given by the three-dimensional Ricci scalar, i.e. the Friedmann equation,
        \begin{equation}\label{eq:3Ricci_scalar_f(Q)}
        \eqalign{
        \fl \, ^{3}\tilde{R} =& \frac{2}{f'}\rho + \frac{f}{f'} - \mathcal{Q} - \frac{2}{3} \tilde{\Theta}^{2} + 2\left(\tilde{\sigma}^{2} - \tilde{\omega}^{2}\right) +\\ 
        \fl &+ 2\frac{f''}{f'} \partial_{h}\mathcal{Q}\left( P^{hi}{}_{i} - h^{ij}P^{h}{}_{ij} - P^{h}{}_{ij} u^{i}u^{j} \right),
        }  
        \end{equation}
        and the projected traceless three-dimensional Ricci tensor,
        \begin{equation}\label{eq:3Ricci_f(Q)}
        \eqalign{
        \fl \bigg( h_{i}{}^{p}h_{j}{}^{q} - \frac{1}{3}h_{ij}h^{pq} \bigg)  \, ^{3}\tilde{R}_{pq} =& \frac{1}{f'} \left[ \pi_{ij} - 2 f'' \partial_{h}\mathcal{Q}\left( h_{i}{}^{p}h_{j}{}^{q} - \frac{1}{3} h_{ij} h^{pq}\right)P^{h}{}_{pq}\right] +\\
        \fl &- \tilde{\Theta}\tilde{\sigma}_{ij} + \tilde{\Theta}\tilde{\omega}_{ij} + \tilde{D}_{\langle i}\mathring{{u}}_{j\rangle} - \tilde{D}_{[i}\mathring{{u}}_{j]}  + \mathring{{u}}_{\langle i}\mathring{{u}}_{j\rangle  } +\\
        \fl &- \mathring{\tilde{\sigma}}_{ij} + \mathring{\tilde{\omega}}_{ij}. 
        }
        \end{equation}
\end{itemize}
As we will see in the following sections, Eqs. \eref{eq:raychaudhuri_equation}-\eref{eq:3Ricci_f(Q)}, together with the energy-momentum conservation law \eref{eq:energy-momentum_conservation} and Eq. \eref{eq:non_metricity_decomposition}, form a closed system able to describe the evolution of Bianchi type-I universes. 


\section{Bianchi type-I model}\label{sec:bianchi_model}

Bianchi type-I models describe anisotropic and homogeneous universes characterized by zero vorticity, $\omega_{ij} = 0$, and autoparallel world lines, $u^{i}\nabla_{i}u^{j} = 0$. In particular, these conditions and the identity $\omega_{ij}=\tilde\omega_{ij}$ imply that the congruence is hypersurface orthogonal. Moreover, in Bianchi type-I models the spatial hypersurfaces foliating the universe are assumed flat, i.e. $\, ^{3}R^{h}{}_{kij} = 0$.

The symmetries of the Bianchi I models have an impact also on the form of the nonmetricity tensor. In fact, remembering that we have chosen $Q_0=0$, we can assume without loss of generality that the only non-zero projections of the nonmetricity tensor are $Q_1$ and $Q^{(0)}{}_{ij}$, so that the nonmetricity tensor results to be of the particular form 
\begin{equation}\label{eq:nonmetricity_decomposition_bianchi}
    Q_{kij} = - \frac{1}{3}Q_{1}u_{k}h_{ij} - Q^{(0)}{}_{ij}u_{k}.
\end{equation}
In local coordinates, expression \eref{eq:nonmetricity_decomposition_bianchi}, and in particular the condition $Q_0=0$, can be justified by adopting the so-called coincidence gauge $\Gamma_{ij}{}^{h}=0$, which is a common assumption in $f(\cal Q)$ gravity. Since the projections of the nonmetricity tensor are tensor quantities, once the identity \eref{eq:nonmetricity_decomposition_bianchi} has been proved in the coincidence gauge, it remains valid in any other gauge. For more detail on this point, the reader is referred to \ref{appendix:bianchi_coordinates}. 

With these assumptions, and separating the Levi-Civita from the nonmetricity contributions, we can write:
\begin{equation}\label{eq:Theta_bianchi_gen}
    \Theta = \tilde{\Theta} - \frac{1}{2}Q_{1} ,
\end{equation}
\begin{equation}\label{eq:sigma_bianchi_gen}
    \sigma_{ij} = \tilde{\sigma}_{ij} - \frac{1}{2}Q^{(0)}{}_{ij} ,
\end{equation}
 and
\begin{equation}\label{eq:acc_bianchi_gen}
    u^{i}\nabla_{i}u^{j}=u^{i}\tilde{\nabla}_{i}u^{j} + N_{ik}{}^{j}u^{k}u^{i}.
\end{equation}
However, as we have seen in Section \ref{sec:f_Q_theory}, in the formulation of $f(\mathcal{Q})$ gravity that we are considering, the Curvature tensor is identically zero, which corresponds to flat spacetime. How can we consider Bianchi I metrics in this setting? The answer is that our assumptions imply that the spacetime manifold is of the type Bianchi type-I {\it with respect to the Levi-Civita connection}. In other words we set $\Theta=0$, $\sigma_{ij}=0$, and $u^{i}\nabla_{i}u^{j}=0$ in such a way that
\begin{equation}\label{eq:Theta_bianchi}
    \tilde{\Theta} = \frac{1}{2}Q_{1} ,
\end{equation}
\begin{equation}\label{eq:sigma_bianchi}
    \tilde{\sigma}_{ij} = \frac{1}{2}Q^{(0)}{}_{ij} ,
\end{equation}
\begin{equation}\label{eq:bianchi_nonmetricity}
    \mathcal{Q} = - \frac{1}{4}Q^{(0)}{}_{ij}Q^{(0)}{}^{ij} + \frac{1}{6}Q_{1}{}^{2} = - 2\tilde{\sigma}^{2} + \frac{2}{3}\tilde{\Theta}^{2},
\end{equation}
and, inserting Eq. \eref{eq:nonmetricity_decomposition_bianchi} into \eref{eq:acc_bianchi_gen} we have
\begin{equation}\label{eq:levi_civita_acceleration_bianchi}
    \mathring{u}^{j} = 0.
\end{equation}
In addition, Eq. \eref{eq:gauss_relation_levi_civita} leads to
\begin{equation}\label{eq:3riemann_levi_civita_bianchi}
    \, ^{3}\tilde{R}^{h}{}_{kij} =0.
\end{equation}
In the subsequent discussion, we consider a matter source described by the energy-momentum tensor
\begin{equation}\label{eq:energy_momentum_tensor_perfect_fluid}
    \Psi_{ij} = \rho u_{i}u_{j} + p h_{ij} + \pi_{ij},
\end{equation}
where $p$ and $\rho$ satisfy the barotropic linear equation of state,
\begin{equation}
p = w\rho, \qquad w=const.
\end{equation}
Inserting all the above results into Eqs. \eref{eq:raychaudhuri_equation}, \eref{eq:3Ricci_scalar_f(Q)} and \eref{eq:3Ricci_f(Q)}, we can write the $1+3$ cosmological equations for Bianchi type-I universes:
\begin{itemize}
\item Raychaudhuri equation,
\begin{equation}\label{eq:raychaudhuri_equation_bianchi}
    \fl \mathring{{\tilde{\Theta}}} + \frac{1}{3}\tilde{\Theta}^{2} + 2 \tilde{\sigma}^{2} + \frac{1}{2f'}\left( \rho + 3 p \right) - \frac{1}{2}\left(\frac{f}{f'} - \mathcal{Q} \right) + \frac{f''}{f'} \tilde{\Theta} \mathring{\mathcal{Q}} = 0;
\end{equation}
\item Spatial equations,
\begin{equation}\label{eq:3R_Bianchi}
    \fl 2 \tilde{\sigma}^{2} - \frac{2}{3}\tilde{\Theta}^{2} +  \frac{2}{f'}\rho + \frac{f}{f'} - \mathcal{Q} = 0,
\end{equation}
\begin{equation}\label{eq:3Ricci_Bianchi}
    \fl \mathring{\tilde{\sigma}} + \tilde{\Theta}\tilde{\sigma} + \frac{f''}{f'}\tilde{\sigma}\mathring{\mathcal{Q}} - \frac{1}{2f'}\frac{\pi_{ij}\tilde{\sigma}^{ij}}{\tilde{\sigma}}= 0;
\end{equation}
\item Energy-momentum conservation, 
\begin{equation}\label{eq:energy_momentum_conservation_bianchi}
    \fl \mathring{\rho} + \tilde{\Theta} \left( \rho + p \right) + \pi^{ij}\tilde{\sigma}_{ij} = 0.
\end{equation}
\end{itemize}
Equation \eref{eq:3Ricci_Bianchi} is obtained multiplying Eq. \eref{eq:3Ricci_f(Q)} by $\tilde{\sigma}^{ij}/\left(2\tilde{\sigma}\right)$, whereas Eq. \eref{eq:energy_momentum_conservation_bianchi} is derived by the temporal projection of Eq. \eref{eq:energy-momentum_conservation}\footnote{We should remark here that the derivatives are with respect to the proper time, not the coordinate one. The two time parameterizations coincide only when $g_{00} = -1$.}.

\section{Dynamical System}\label{sec:dynamical_systems}
In this section, we will apply the DSA to analyze the dynamics of Bianchi type-I universes in the framework of $f(\mathcal{Q})$ gravity. We will deal with some specific models associated with particular functions $f(\mathcal{Q})$, all widely used in literature. In one of the examples, we will also consider the presence of anisotropic pressure. 

In our analysis we will always consider an expanding universe, hence $\tilde{\Theta}>0$.

\subsection{\texorpdfstring{$f(\mathcal{Q})$}{} as a power law without anisotropic pressure}\label{sec:alphaQ}
As a first example, we consider the function 
\begin{equation}\label{eq:f(Q)_example_1}
    f(\mathcal{Q})=\alpha\mathcal{Q}^{n},
\end{equation}
with $\alpha$ a dimensional constant, and a null anisotropic pressure $\pi_{ij} = 0$. In this case, Eqs. \eref{eq:raychaudhuri_equation_bianchi}-\eref{eq:energy_momentum_conservation_bianchi} assume the form,
\begin{equation}\label{eq:raychaudhuri_equation_bianchi_alphaQ}
    \fl \mathring{\tilde{\Theta}} + \frac{1}{3}\tilde{\Theta}^{2} + 2 \tilde{\sigma}^{2} + \frac{n-1}{2n}\mathcal{Q} + \left(n-1\right)\tilde{\Theta}\frac{\mathring{\mathcal{Q}}}{\mathcal{Q}} +\frac{1}{2 \alpha  n}\left(1 + 3w \right) \mathcal{Q}^{1-n} \rho  = 0,
\end{equation}
\begin{equation}\label{eq:3R_Bianchi_alphaQ}
   \fl 2\tilde{\sigma}^{2}  - \frac{2}{3}\tilde{\Theta}^{2} + \frac{1-n}{n} \mathcal{Q} 
   +\frac{2}{\alpha  n}\mathcal{Q}^{1-n}\rho  = 0,
\end{equation}
\begin{equation}\label{eq:3Ricci_Bianchi_alphaQ}
    \fl \mathring{\tilde{\sigma}}+ \tilde{\Theta}\tilde{\sigma} + \left(n-1\right) \tilde{\sigma} \frac{\mathring{\mathcal{Q}}}{\mathcal{Q}} = 0,
\end{equation}
\begin{equation}\label{eq:energy_momentum_alphaQ}
    \fl \mathring{\rho} + \tilde{\Theta} \left( 1 + w \right)\rho = 0.
\end{equation}
In order to recast these equations in a form more suitable for a dynamical system analysis, we define the following dimensionless variables \footnote{In the natural units we are using, the quantities $\Theta^{2}$, $\sigma^{2}$, $\rho$, and $\mathcal{Q}$ have the dimension of a length to the power of $-2$.},
\begin{equation}\label{eq:dynamical_variables_1}
    \Sigma^{2} = 3 \frac{\tilde{\sigma}^{2}}{\tilde{\Theta}^{2}}, \qquad 
    \Omega^{2} = 3 \frac{1}{\alpha}\frac{1}{\tilde{\Theta}^{2n}}\rho.
\end{equation}
Notice that the dynamical variables related to the shear and the matter sources have been chosen non-negative, offering the advantage of a partial compactification of the phase space. The choice of the matter variable should also be discussed. In general, one chooses as the variable associated to $\rho$, simply $3\rho^{2}/\tilde{\Theta}^{2}$ or $3\rho^{2}/(f' \tilde{\Theta}^{2})$, which directly relates to the cosmic matter parameters, and therefore, it is easier to compare with observational results. In fact these  parameters appear via Eq. \eref{eq:3R_Bianchi} in many observable quantities, such as the luminosity distance relation of the lookback time etc.  
However, here and in the following examples, except for Sec. \ref{sec:expQ}, we choose a different form for $\Omega$. The reason is that such form allows us to introduce fewer dynamical variables. In addition, our choice allows us to obtain a variable that involves only the expansion rate and energy density, which can be measured independently, leading to an equally good variable in terms of comparison with observations.

We also define the ``average length scale'' $l$ by using $\tilde{\Theta}$,
\begin{equation}
    \frac{\mathring{l}}{l} = \frac{1}{3}\tilde{\Theta},
\end{equation}
that allows us to introduce the conformal time,
\begin{equation}
    \mathcal{T} = \hbox{ln} \: l.
\end{equation}
Making use of the above variables and recalling the identity \eref{eq:bianchi_nonmetricity},
we can rewrite the system of cosmological equation in the new form,
\begin{equation}\label{eq:dynamical_equations_1_2}
    \fl (1 - 2n) \left(1 - \Sigma^{2}\right) + \left(\frac{3}{2}\right)^{n-1} \Omega^{2} \left(1 - \Sigma^{2} \right)^{1-n} = 0,
\end{equation}
\begin{equation}\label{eq:dynamical_equations_1_3}
    \fl \frac{\hbox{d}\Sigma}{\hbox{d}\mathcal{T}} = \frac{1}{3n}\Sigma  \left(\Sigma^{2} - 1\right) \left[3 (n+1) - 3^{n} (3 w+1) \left(2-2 \Sigma ^2\right)^{-n} \Omega^{2}\right].
\end{equation}
\begin{table}[t]
\centering
\renewcommand{\arraystretch}{1.5}
\caption{The stability of the fixed points and evolution of $l$, $\tilde{\sigma}$, and $\rho$ for $f(\mathcal{Q})=\alpha\mathcal{Q}^{n}$ and $\pi_{ij}
=0$. The parameters $\tau_{0}$, $l_{0}$, $\sigma_{0}$, $\sigma_{1}$, $\rho_{0}$, and $\rho_{1}$ are constants of integration.}
\begin{tabular}{ lccccccc }
\toprule
 &  \multicolumn{3}{c}{$w=0$} & &  \multicolumn{3}{c}{$0 < w \leq 1$} \\
\cmidrule{2-4} \cmidrule{6-8}
Point &  Attractor & Repeller & Saddle & & Attractor & Repeller & Saddle \\
\midrule
$P_{1}$ &  $n\geq \frac{1}{2}$ & & & & $\frac{1}{2} \leq n < \frac{1 + w}{2w}$ & $n > \frac{1 + w}{2w}$ & \\
\toprule
&  \multicolumn{2}{c}{Average length} & \multicolumn{2}{c}{Shear} & \multicolumn{2}{c}{Energy density}\\
\midrule
$P_{1}$ &  \multicolumn{2}{c}{$l = l_{0} \left( \tau - \tau_{0} \right)^{\frac{2 n}{3 (1+w)}}$} & \multicolumn{2}{c}{$\tilde{\sigma} = \sigma_{0} = 0$} & \multicolumn{2}{c}{$\rho = \rho_{0} + \frac{\rho_{1}}{\left( \tau - \tau_{0} \right)^{2n}}$}\\
\bottomrule
\end{tabular}
\label{table_1}
\end{table}
From Eq. \eref{eq:dynamical_equations_1_2} we derive $\Omega$ as a function of $\Sigma$,
\begin{equation}\label{Omega}
    \Omega = \sqrt{2n-1}\left(\frac{3}{2}\right)^{\frac{1-n}{2}}  \left(1 - \Sigma^{2}\right)^{\frac{n}{2}},
\end{equation}
which makes Eq. \eref{eq:dynamical_equations_1_3} a differential equation for $\Sigma$,
\begin{equation}\label{eq:dynamical_equations_1_4}
   \frac{\hbox{d}\Sigma}{\hbox{d}\mathcal{T}} = \frac{3}{2n} \Sigma  \left(1 - \Sigma^{2}\right) \left[(2 n-1) w-1\right].
\end{equation}
A first consideration about the above equations is that $\Sigma = 1$ is not an acceptable value, since we derive Eq. \eref{Omega} from Eq. \eref{eq:dynamical_equations_1_2}, assuming that $\Sigma\neq1$. The same problem will occur in Sec. \ref{sec:alphaQ_anisotropic_pressure}, in which the function $f(\mathcal{Q})$ is again \eref{eq:f(Q)_example_1}.

Furthermore, being $\Omega$ and $\Sigma$ non-negative, Eq. \eref{Omega} is only meaningful if $n \geq 1/2$ and $0 \leq \Sigma < 1$, and for some values of $n$ in the intervals $n \geq 1/2$ and $\Sigma > 1$, or $n \leq 1/2$ and $\Sigma > 1$. However, we consider only the condition $0 \leq \Sigma < 1$. This choice has two motivations. The first is that, from a physical point of view, we are interested in the states of phase space describing an isotropic universe, i.e. $\Sigma=0$. This state cannot be reached by any orbit starting at $\Sigma > 1$. A second motivation is that in Eq. \eref{Omega} there is the term  $\left(1 - \Sigma^{2}\right)^{\frac{n}{2}}$, the value of which depends strictly on the choice of $n$ (e.g. even, odd, or a rational number) when $\Sigma > 1$. The case $n \geq 1/2$ and $0 \leq \Sigma \leq 1$, on the other hand, being a continuous interval for $n$, offers a wider setting for a parameter analysis aimed to comparison with observations. Similar constraints will be necessary also in the models we will consider in the following sections.

\begin{figure}[t]
\centering
    \begin{subfigure}[ht]{0.49\textwidth}
        \includegraphics[width=\linewidth]{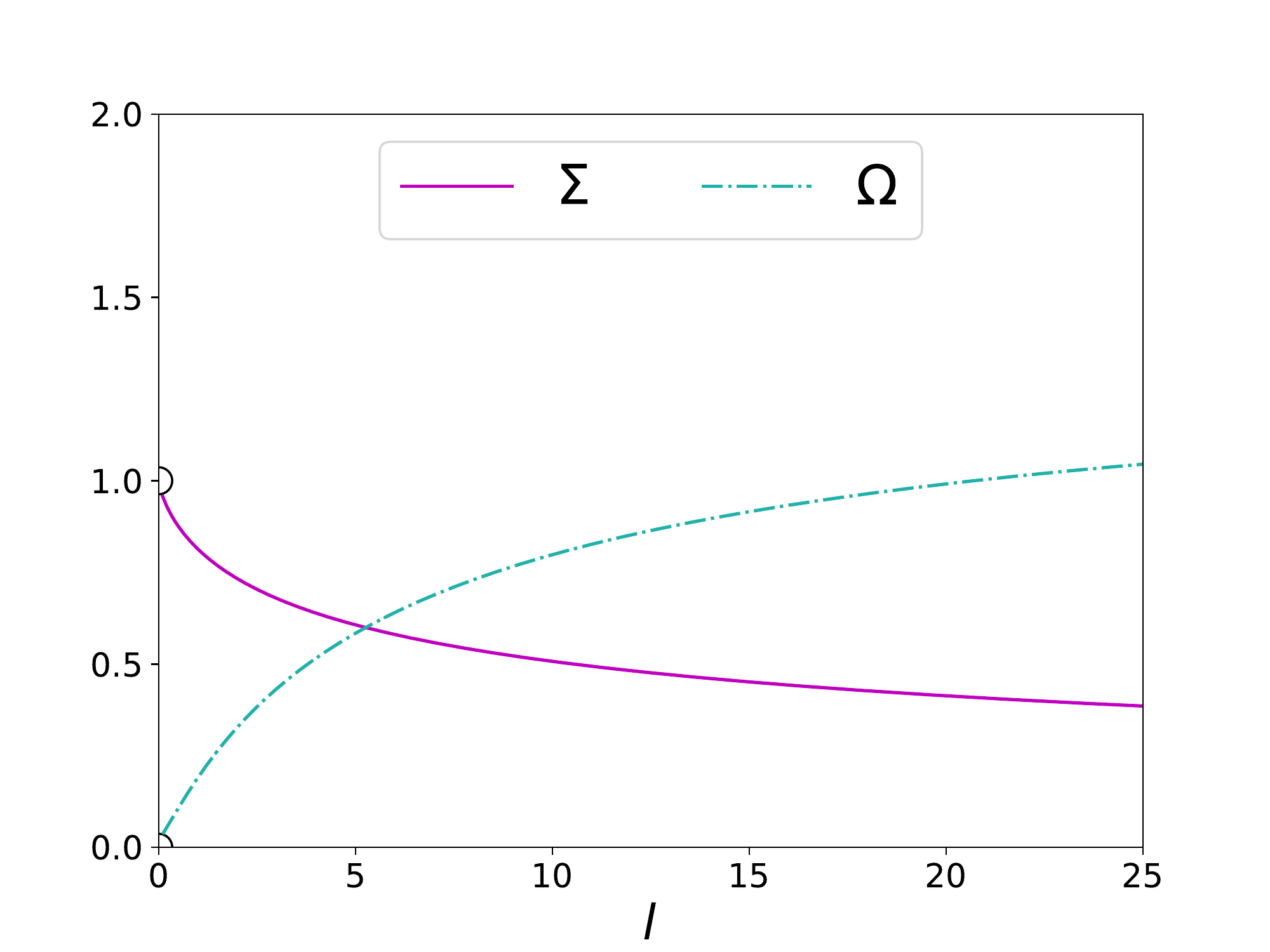}
        \caption{}
        \label{fig:Phase_space_alpha_Q_n_0_1}
    \end{subfigure}
    \hfill
    \begin{subfigure}[ht]{0.49\textwidth}
        \includegraphics[width=\linewidth]{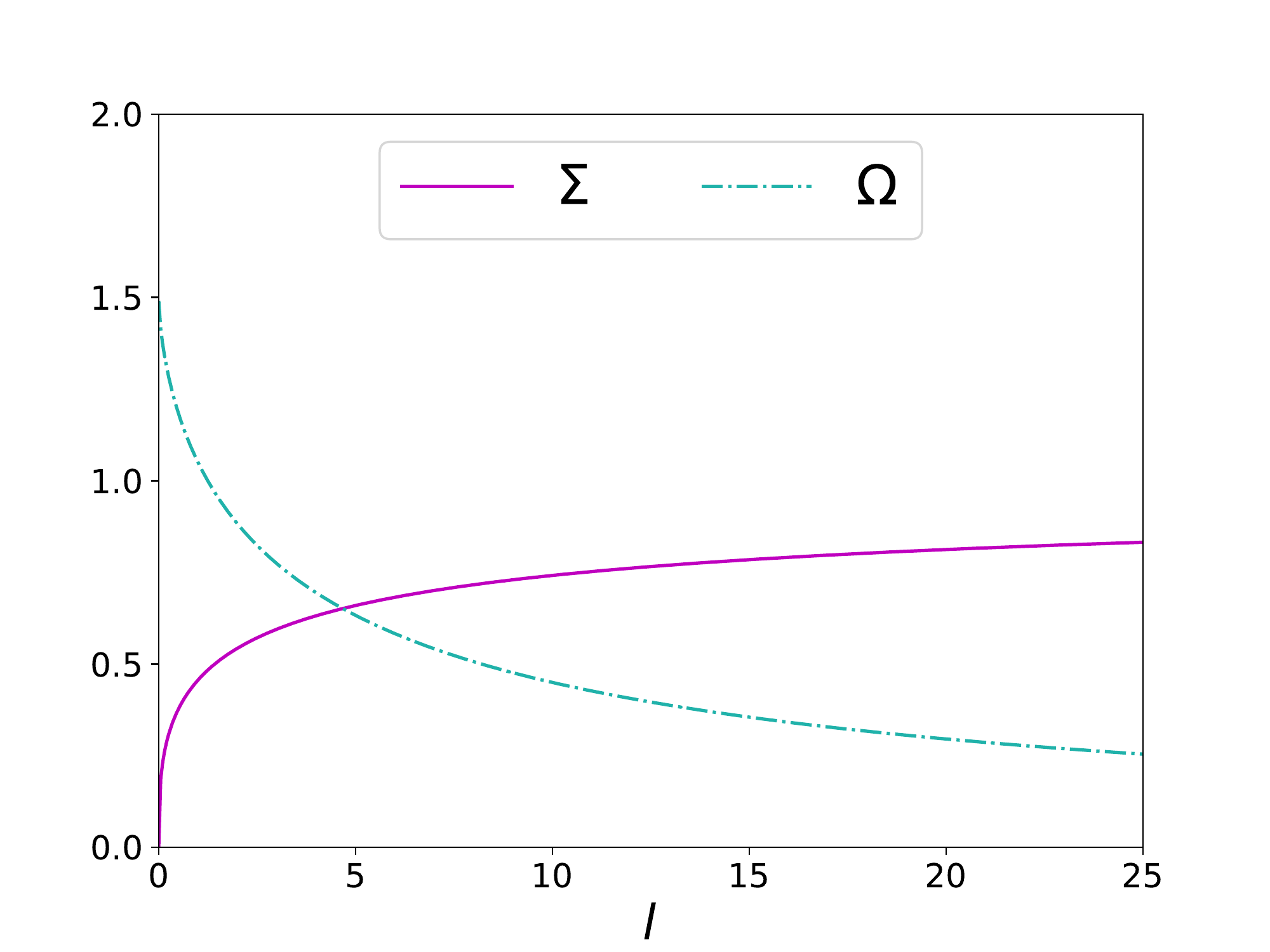} 
        \caption{}
        \label{fig:Phase_space_alpha_Q_n_0_2}
    \end{subfigure}
    \hfill
    \begin{subfigure}[ht]{0.49\textwidth}
        \includegraphics[width=\linewidth]{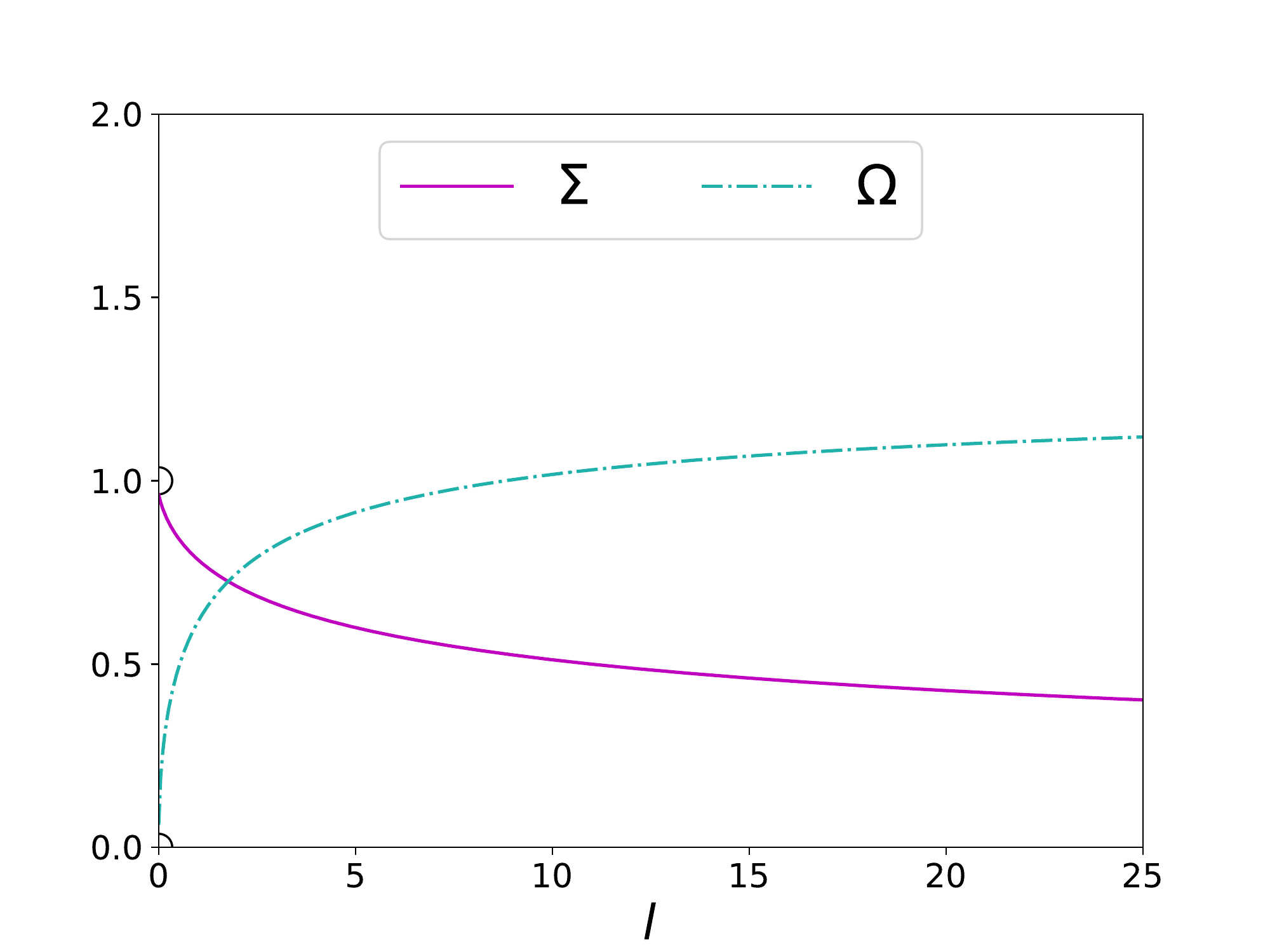} 
        \caption{}
        \label{fig:Phase_space_alpha_Q_n_0_3_ver_3}
    \end{subfigure}
    \caption{Evolution of $\Sigma(l)$ and $\Omega(l)$ with: (a) $w=0$, $n=3$, and $\Sigma_{1}=-1/3$; (b) $w=1/3$, $n=3$, and $\Sigma_{1}=2/3$; (c) $w=1/3$, $n=3/2$, and $\Sigma_{1}=-1/4$. The empty (half-)circles represent the conditions $\Sigma \neq 1$ and $\Omega \neq 0$.}
    \label{fig:Phase_space_alpha_Q_n_0}
\end{figure}

The  system \eref{Omega} and \eref{eq:dynamical_equations_1_4} presents only one critical point,
\begin{equation}
    P_{1} = \Bigg\lbrace \Sigma = 0,\: \Omega = \sqrt{2n-1}\left( \frac{3}{2} \right)^{\frac{1-n}{2}} 
    \Bigg\rbrace,
\end{equation}
which represents a universe where the shear is negligible with respect to the matter.

The derivative of Eq. \eref{eq:dynamical_equations_1_4} with respect to $\Sigma$ allows us to discuss the stability of the solutions near the critical point, which depends on the values of $w$ and $n$. The results are shown in Table \ref{table_1}.

We can obtain an ``approximation'' for the time dependence of $l$, $\tilde{\sigma}$, and $\rho$ near to a critical point by substituting Eq. \eref{eq:dynamical_variables_1} into Eqs. \eref{eq:raychaudhuri_equation_bianchi_alphaQ}, \eref{eq:3Ricci_Bianchi_alphaQ}, and \eref{eq:energy_momentum_alphaQ}. The results are again reported in Table \ref{table_1}.

Equation \eref{eq:dynamical_equations_1_4} can be solved analytically, so we can also obtain exact solutions for $\Sigma$ and $\Omega$ as a function of average length scale $l$,
\begin{equation}\label{eq:SolEs1}
\eqalign{
    \Sigma(l) &= \frac{1}{\sqrt{1+e^{2 \Sigma_{1}} l^{\frac{3 (1 + w - 2nw)}{n}}}},\\
    \Omega(l) &= \sqrt{2n-1}\left(\frac{3}{2}\right)^{\frac{1-n}{2}} \left(\frac{e^{2 \Sigma_{1}} l^{\frac{3 (1 + w - 2nw)}{n}}}{1+e^{2 \Sigma_{1}} l^{\frac{3 (1 + w - 2nw)}{n}}}\right)^{n/2},
    }
\end{equation}
where $\Sigma_{1}$ is a constant of integration. Using Eq. \eref{eq:SolEs1}, we can compare the evolution of $\Sigma$ and $\Omega$ with the results coming from the stability analysis in Table \ref{table_1}. As it can be seen in Figure \ref{fig:Phase_space_alpha_Q_n_0}, once the appropriate parameters have been chosen, the results are consistent with Table \ref{table_1}. In Figure \ref{fig:Phase_space_alpha_Q_n_0_1}, where $w=0$ and $n \geq 1/2$, we have that $P_{1}$ is an attractor, whereas in Figure \ref{fig:Phase_space_alpha_Q_n_0_2}, and Figure \ref{fig:Phase_space_alpha_Q_n_0_3_ver_3}, with $0 < w \leq 1$, $P_{1}$ is a repeller or an attractor, respectively, depending on the value of $n$.

\begin{figure}[t]
\centering
\begin{subfigure}[ht]{0.49\textwidth}
    \includegraphics[width=\linewidth]{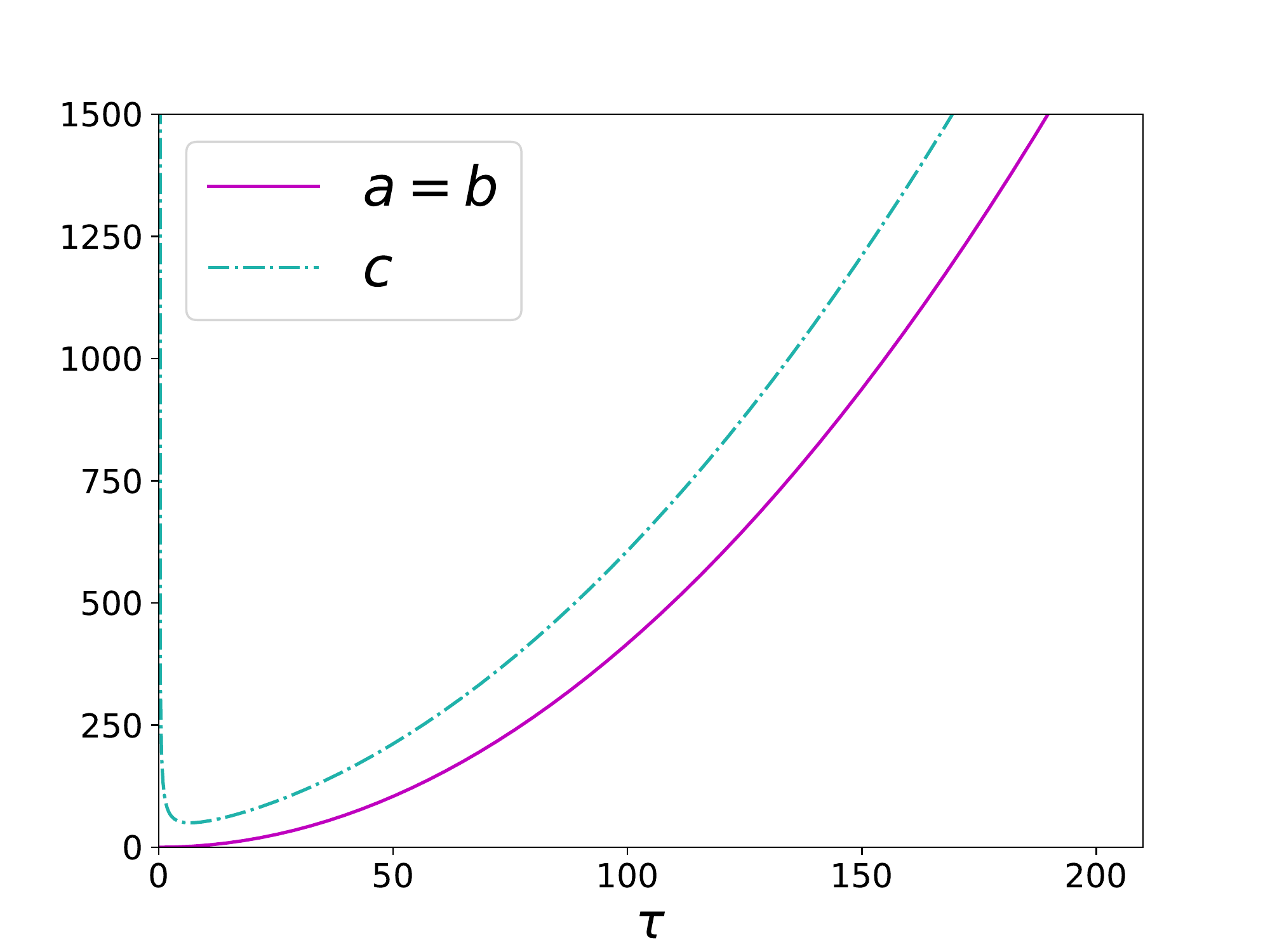}
    \caption{}
    \label{fig:comparison_1}
\end{subfigure}
\begin{subfigure}[ht]{0.49\textwidth}
    \includegraphics[width=\linewidth]{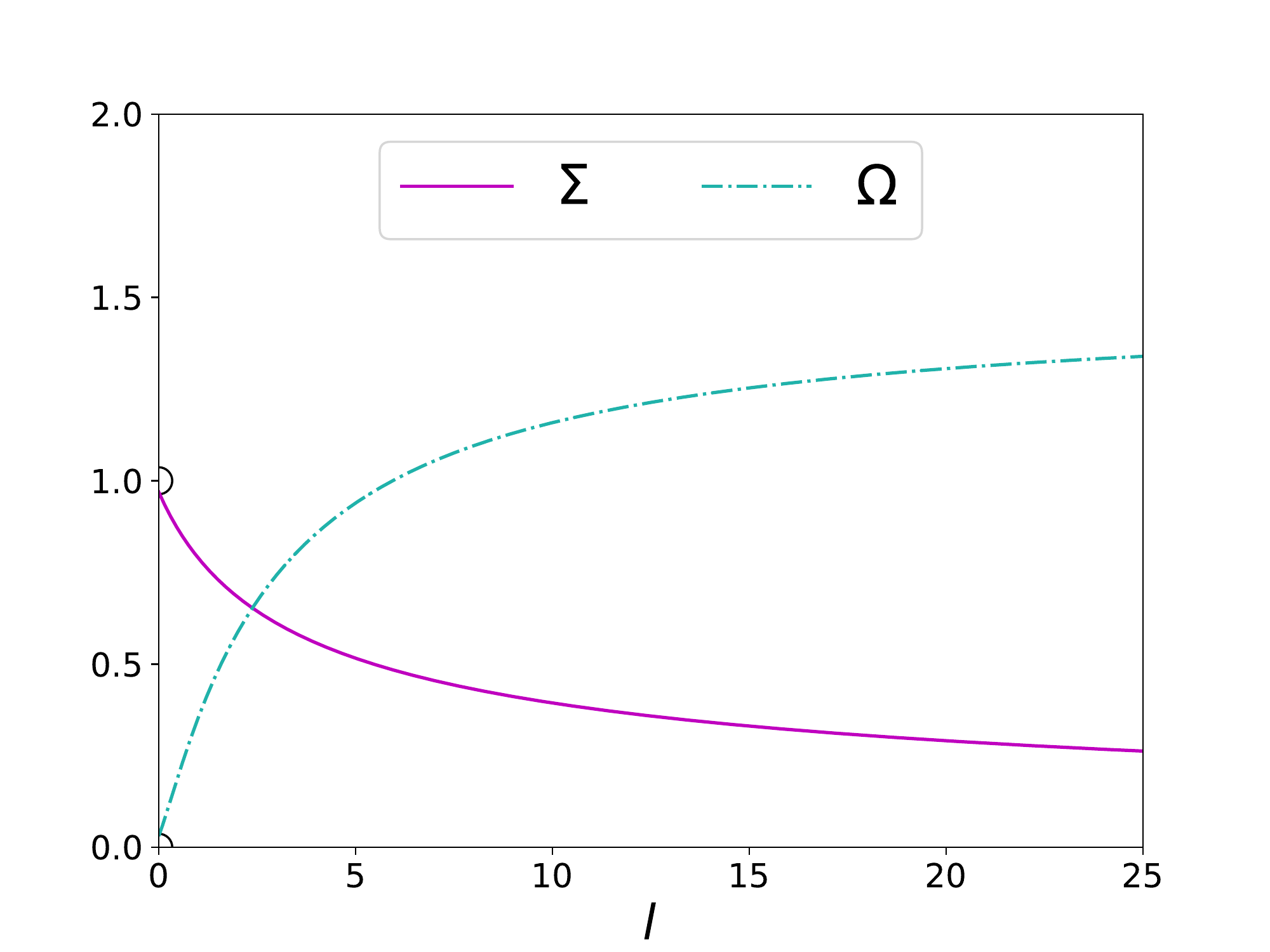}
    \caption{}
    \label{fig:comparison_2}
\end{subfigure}
\caption{(a) Evolution of the scale factor $a$, $b$, and $c$ in function of the proper time $\tau$, with $w=0$ and $n=3$. (b) Evolution of $\Sigma$, and $\Omega$ in function of the average length scale $l$, with $w=0$ and $n=3$. The empty (half-)circles represent the conditions $\Sigma \neq 1$ and $\Omega \neq 0$.}
\label{fig:comparison}
\end{figure}

In \cite{Esposito:2021ect}, the reconstruction method was used to find exact Bianchi type-I cosmologies in $f(\mathcal{Q})$ gravity. It is interesting to compare these results with the more general description we have obtained from the above phase space analysis.

For example, in \cite{Esposito:2021ect} it was found, for $f(\mathcal{Q})=\mathcal{Q}^{n}$, $w=0$ and $n$ an odd integer, the following solution for the scale factors,
\begin{equation}\label{eq:scale_factors}
\eqalign{
    a(\tau) =& a_{1} \left[(\tau-\tau_{0})^2 - K^{2} \right]^{\frac{n}{3}}\left[ \frac{\left( \tau - \tau_{0} \right) -  K}{\left( \tau - \tau_{0} \right) +  K} \right]^{\frac{n}{3}},\\
    c(\tau) =& c_{1} \left[(\tau-\tau_{0})^2 - K^{2}\right]^{\frac{n}{3}} \left[\frac{\left( \tau - \tau_{0} \right) - K}{\left( \tau - \tau_{0} \right) + K} \right]^{-\frac{2n}{3}},
    }
\end{equation}
represented by Figure \ref{fig:comparison_1}. $K$ is a parameter of the theory, while $a_{1}$, $c_{1}$, and $\tau_{0}$ are constants of integration. It is clear that the scale factors tend to have the same expansion rate as the time increases, thus describing a universe which tends to isotropize. Such isotropization is evident in Figure \ref{fig:comparison_2}, which shows the behavior of $\Sigma$ and $\Omega$ calculated for the \eref{eq:scale_factors}. As expected, this behavior matches exactly the one of Figure \ref{fig:Phase_space_alpha_Q_n_0_1} once the parameter are chosen in a consistent way.

\subsection{\texorpdfstring{$f(\mathcal{Q})$}{} as a power law with anisotropic pressure}\label{sec:alphaQ_anisotropic_pressure}

We consider again the function $f(\mathcal{Q}) = \alpha\mathcal{Q}^{n}$, but now we add an anisotropic pressure of the form (see e.g. \cite{ellis_maartens_maccallum_2012} and \cite{PhysRev.58.919}),
\begin{equation}
    \pi_{ij} = - \mu \tilde{\sigma}_{ij} ,
\end{equation}
being $\mu$ a suitable dimensional constant.

Under these assumptions, the dynamical equations are:
\begin{equation}\label{eq:raychaudhuri_equation_bianchi_alphaQ_pi}
    \fl \mathring{\tilde{\Theta}} + \frac{1}{3}\tilde{\Theta}^{2} + 2 \tilde{\sigma}^{2} + \frac{n-1}{2n}\mathcal{Q} + \left(n-1\right)\tilde{\Theta}\frac{\mathring{\mathcal{Q}}}{Q} + \frac{1}{2 \alpha  n}\left(1 + 3w \right) Q^{1-n} \rho  = 0,
\end{equation}
\begin{equation}\label{eq:3R_Bianchi_alphaQ_pi}
   \fl 2\tilde{\sigma}^{2}  - \frac{2}{3}\tilde{\Theta}^{2} + \frac{1-n}{n} \mathcal{Q} 
   +\frac{2}{\alpha  n}\mathcal{Q}^{1-n}\rho  = 0,
\end{equation}
\begin{equation}\label{eq:3Ricci_Bianchi_alphaQ_pi}
    \fl  \mathring{\tilde{\sigma}}+ \tilde{\Theta}\tilde{\sigma} + \left(n-1\right) \tilde{\sigma} \frac{\mathring{\mathcal{Q}}}{Q} + \frac{\mu }{\alpha  n} \mathcal{Q}^{1-n}\tilde{\sigma} = 0,
\end{equation}
\begin{equation}
    \fl \mathring{\rho} + \tilde{\Theta} \left( 1 + w \right)\rho - 2\mu \sigma^{2} = 0.
\end{equation}
In this case, we have an additional variable related to the anisotropic pressure,
\begin{equation}
    \mathcal{M} = \frac{\mu}{\alpha} \tilde{\Theta}^{1-2n},
\end{equation}
together with
\begin{equation}\label{eq:dynamical_variables_2_2}
    \Sigma^{2} = 3 \frac{\tilde{\sigma}^{2}}{\tilde{\Theta}^{2}}, \quad
    \Omega^{2} = 3 \frac{1}{\alpha}\frac{1}{\tilde{\Theta}^{2n}}\rho.
\end{equation}
By following a similar procedure as in the previous example, we obtain the final system of dynamical equations:
\begin{eqnarray}
    \fl \:\:\:\Omega &=& \sqrt{2n - 1} \left(\frac{3}{2}\right)^{\frac{1-n}{2}} \left(1 - \Sigma^{2}\right)^{\frac{n}{2}},\\
    \fl \frac{\hbox{d}\Sigma}{\hbox{d}\mathcal{T}} &=& -\frac{1}{2n}\left(\frac{3}{2}\right)^{n} \Sigma  \left(1-\Sigma^{2}\right)^{1-n} \left[4 \mathcal{M} + 3^{1-n} \left(2-2 \Sigma^{2}\right)^n (1 + w - 2nw)\right],\label{eq:dynamical_equation_Sigma_prime_2}\\
    \fl \frac{\hbox{d}\mathcal{M}}{\hbox{d}\mathcal{T}} &=& \: \frac{3^{n}}{2n}\mathcal{M}  \Big\{ 8\left(n-1\right) \mathcal{M}  \Sigma ^{2} \left(2-2 \Sigma ^2\right)^{-n} +\nonumber\\
    \fl &&- 3^{1-n} (2 n-1) \left[\Sigma ^{2} \left(2 n w-w-1\right)-w-1\right]\Big\}.\label{eq:dynamical_equation_emme_prime_2}
\end{eqnarray}
As anticipated in Sec. \ref{sec:alphaQ}, in the above system of equations, $\Sigma=1$ is not an acceptable value. In addition, the conditions to have $\Omega$ and $\Sigma$ real and non-negative are $n \geq 1/2$ and $0 \leq \Sigma < 1$.

\begin{table}
\centering
\renewcommand{\arraystretch}{1.5}
\caption{The stability of the fixed points and evolution of $l$, $\tilde{\sigma}$, and $\rho$ for $f(\mathcal{Q})=\alpha\mathcal{Q}^{n}$ and $\pi_{ij}=-\mu\sigma_{ij}$. The parameters $\tau_{0}$, $l_{0}$, $\sigma_{0}$, $\rho_{0}$, and $\rho_{1}$ are constants of integration.}
\begin{tabular}{ lccccccc }
\toprule
 &  \multicolumn{3}{c}{$w=0$} & &  \multicolumn{3}{c}{$0 < w \leq 1$} \\
\cmidrule{2-4} \cmidrule{6-8}
Point &  Attractor & Repeller & Saddle & & Attractor & Repeller & Saddle \\
\midrule
$P_{1}$ & & & $n \geq \frac{1}{2}$ & & & $n>\frac{w+1}{2 w}$ & $\frac{1}{2} \leq n<\frac{w+1}{2 w}$  \\
\toprule
 &  \multicolumn{2}{c}{Average length} & \multicolumn{2}{c}{Shear} & \multicolumn{2}{c}{Energy density}\\
\midrule
$P_{1}$ & \multicolumn{2}{c}{$l = l_{0} \left( \tau - \tau_{0} \right)^{\frac{2 n}{3 (1+w)}}$} &  \multicolumn{2}{c}{$\tilde{\sigma} = \sigma_{0} = 0$} &  \multicolumn{2}{c}{$\rho = \rho_{0} + \frac{\rho_{1}}{\left( \tau - \tau_{0} \right)^{2n}}$}\\
\bottomrule
\end{tabular}
\label{table_2}
\end{table}

The invariant submanifolds of the system include $\Sigma = 0$ and $\mathcal{M}=0$. Notice that the invariant submanifold $\Sigma=0$ represents isotropic universes, whereas $\mathcal{M}=0$ implies that either we are in a situation in which the terms associated to the coupling $\mu$ are negligible (and therefore the universe described in Sec. \ref{sec:alphaQ}) or that the expansion is going to infinity. It is not immediate in this framework to distinguish these two cases, however. Only a more detailed analysis of the equations, or a different choice of variables might shed clarity on this point. We will not attempt such analysis here.

In the parameter range we consider, there is one critical point,
\begin{equation}
    P_{1} = \bigg\{  \Sigma = 0,\: \mathcal{M} = 0, \: \Omega = \sqrt{2 n-1}\left(\frac{3}{2}\right)^{\frac{1-n}{2}}  \bigg\},
\end{equation}
where matter dominates over the shear and the anisotropic pressure. The analysis of the stability and the approximate evolution of $l$, $\tilde{\sigma}$ and $\rho$ are summarized in Table \ref{table_2}.

\begin{figure}[t]
\centering
    \begin{subfigure}[ht]{0.45\textwidth}
        \includegraphics[width=\linewidth]{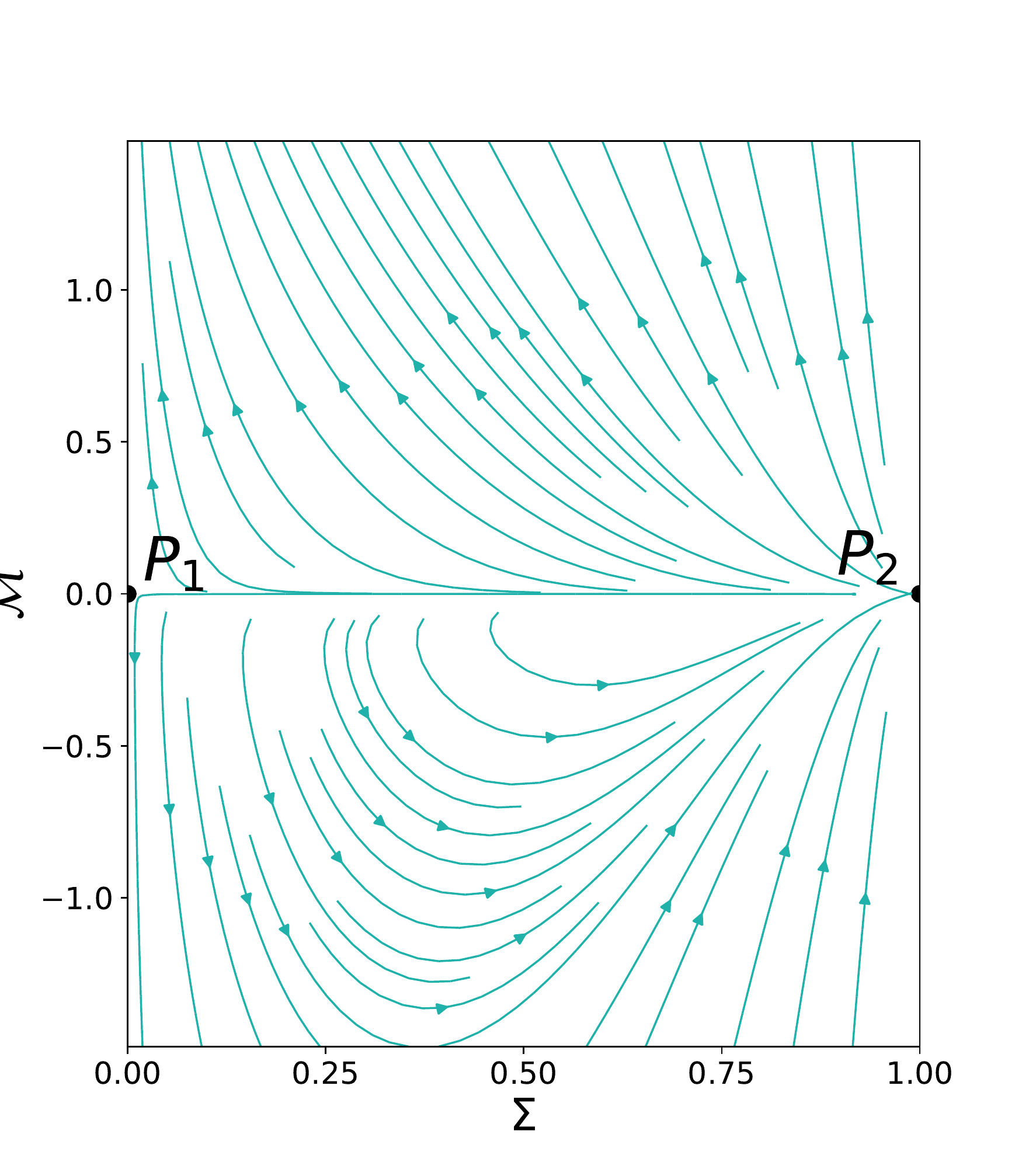} 
        \caption{}
        \label{fig:example2_1}
    \end{subfigure}
    \begin{subfigure}[ht]{0.45\textwidth}
        \includegraphics[width=\linewidth]{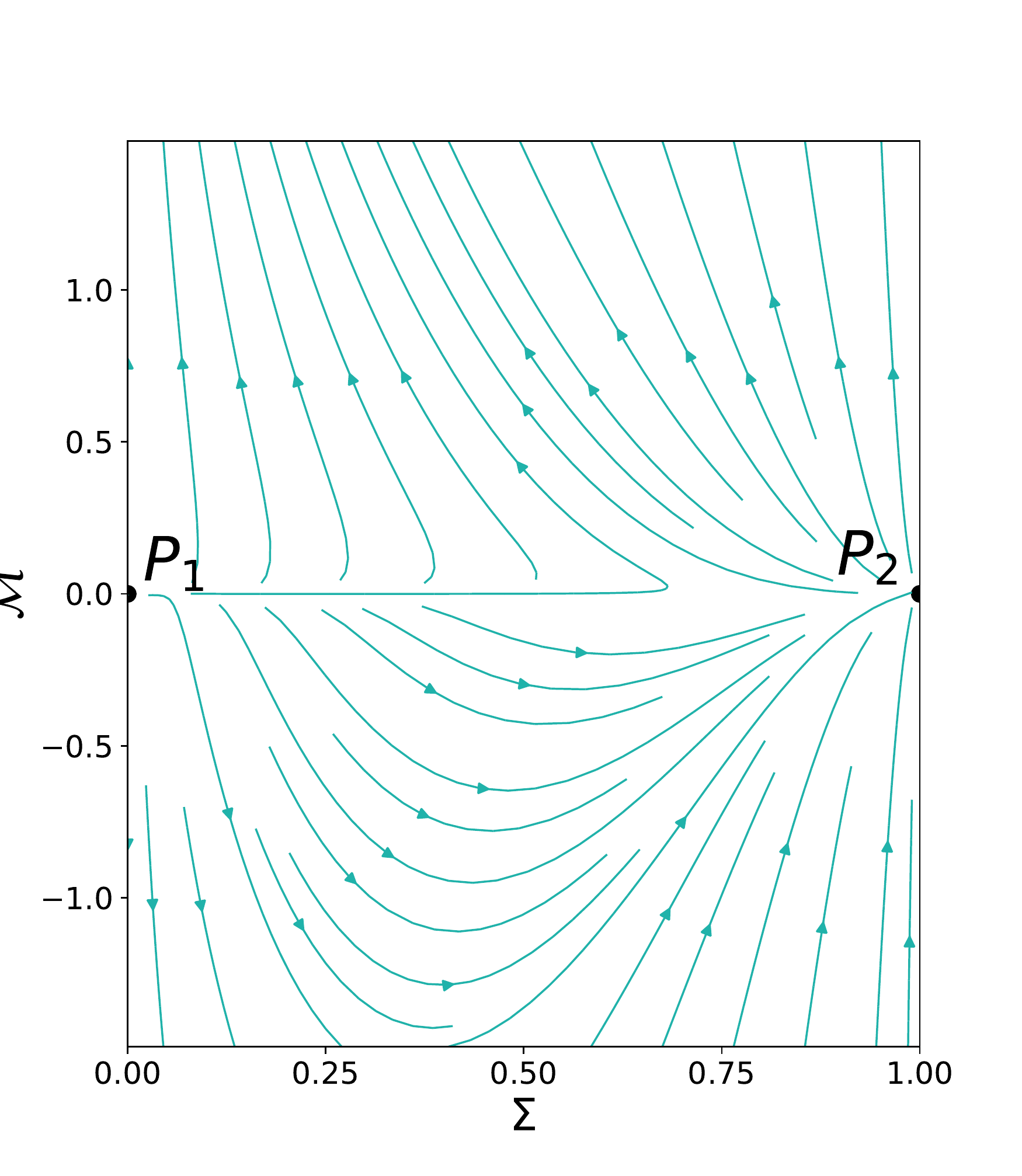} 
        \caption{}
        \label{fig:example2_2}
    \end{subfigure}
    \begin{subfigure}[ht]{0.45\textwidth}
        \includegraphics[width=\linewidth]{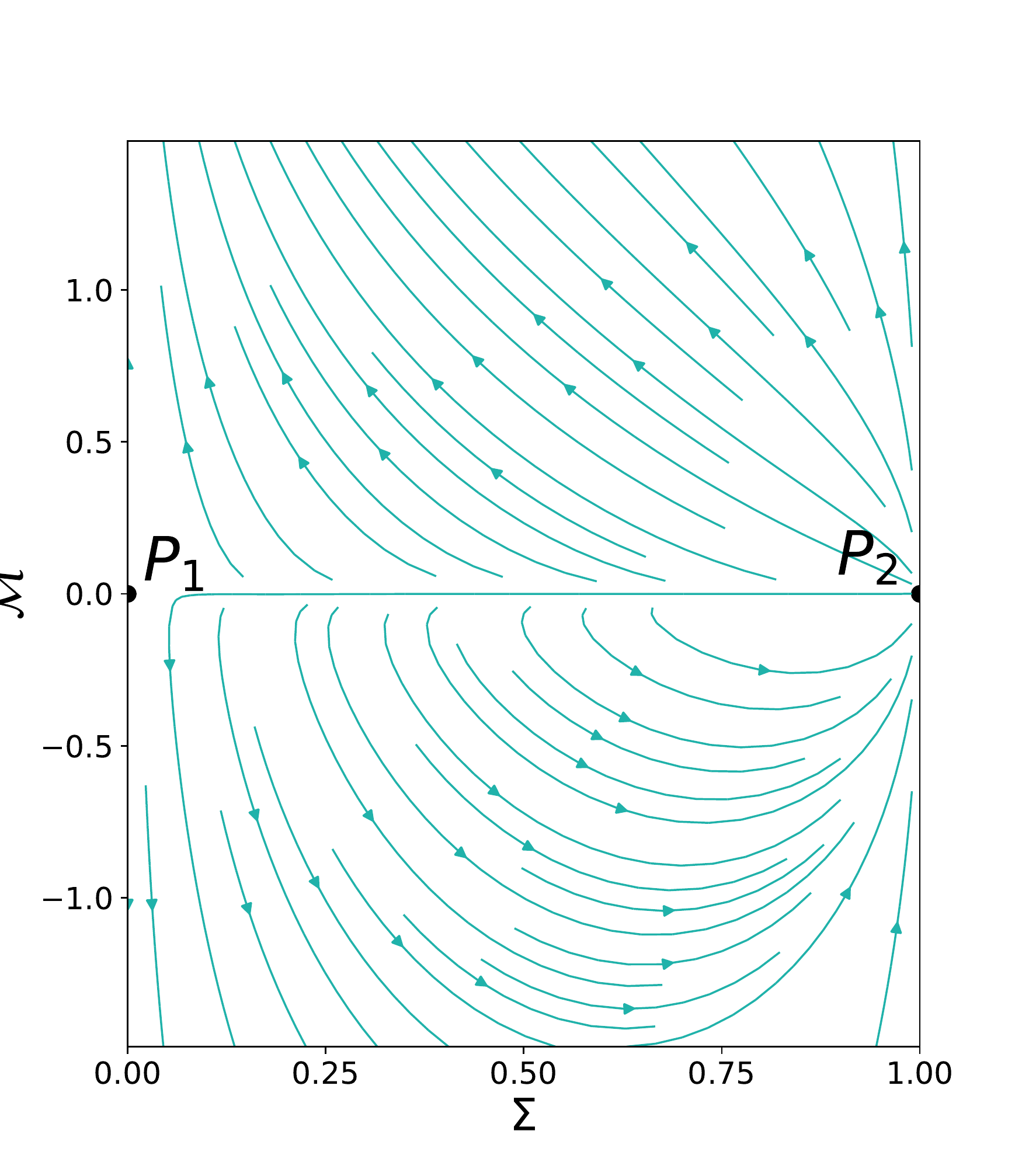}
        \caption{}
        \label{fig:example2_3}
    \end{subfigure}
    \caption{Phase space portrait of the system \eref{eq:dynamical_equation_Sigma_prime_2}-\eref{eq:dynamical_equation_emme_prime_2} with (a) $w=0$ and $n=3$, (b) $w=\frac{1}{3}$ and $n=3$, (c) $w=\frac{1}{3}$ and $n=3/2$.}
    \label{fig:example2}
\end{figure}

The phase space is described in Figures \ref{fig:example2_1}, \ref{fig:example2_2} and \ref{fig:example2_3}, for different values of $w$ and $n$. To proceed in the analysis, we define
\begin{equation}
    P_{2} := \bigg\{  \Sigma = 1,\: \mathcal{M} = 0, \: \Omega = 0 \bigg\},
\end{equation}
which is {\it not} a critical point, but it will be useful to describe the orbits of the phase space. 

The phase space we obtained shows several types of cosmic evolutions. For example in Figure \ref{fig:example2_1}, close to $P_{2}$, with $\mathcal{M}$ positive, we are in a universe where matter is negligible compared to shear. As the time progresses, the universe isotropizes with a decreasing expansion rate. In contrast, in the negative half-plane for $M$, after a phase of isotropization and approach to $P_{1}$, the orbits return to their starting point, i.e. to an anisotropic state. A similar behavior is found in Figure \ref{fig:example2_3}. On the other hand, in Figure \ref{fig:example2_2} we have that orbits move away from an isotropic universe, represented by the points of the phase space near $P_{1}$. In the positive half-plane, the region near $P_{2}$ is a transition phase for the system leading to a decelerated isotropization, whereas in the negative half-plane, there are decelerated and accelerated expansion phases that lead the universe to anisotropy.

As expected, the invariant submanifold $\mathcal M=0$ mirror exactly the phase space of the case of Section \ref{sec:alphaQ}.

\subsection{The case \texorpdfstring{ $f(\mathcal{Q})=\alpha\left(\sqrt{\mathcal{Q}} + \beta \mathcal{Q}^{n}\right)$}{}}\label{sec:sqrtQ}

We now consider the following function,
\begin{equation}
    f\left(\mathcal{Q}\right) = \alpha\left(\sqrt{\mathcal{Q}} + \beta \mathcal{Q}^{n}\right),
\end{equation}
where $\alpha$ and $\beta$ are dimensional constants, and set the anisotropic pressure $\pi_{ij}$ equal to zero.
The resulting cosmological equations are,
\begin{equation}\label{eq:raychaudhuri_equation_bianchi_alphasqrtQ}
\fl \eqalign{
    \mathring{\tilde{\Theta}} + \frac{1}{3}&\tilde{\Theta}^{2} + 2 \tilde{\sigma}^{2} + \frac{\mathcal{Q}}{2} - \frac{\mathcal{Q} + \beta  \mathcal{Q}^{n+\frac{1}{2}}}{1 + 2 \beta  n \mathcal{Q}^{n-\frac{1}{2}}} +\\
    &- \frac{1}{2} \frac{\mathring{\mathcal{Q}}}{\mathcal{Q}}\frac{1 - 4 \beta  (n-1) n \mathcal{Q}^{n-\frac{1}{2}}}{1 +  2 \beta  n \mathcal{Q}^{n-\frac{1}{2}}}\tilde{\Theta} + \frac{ \sqrt{\mathcal{Q}} }{\alpha  \left(1 + 2 \beta  n Q^{n-\frac{1}{2}}\right)}\left(1 +3 w\right)\rho = 0,
    }  
\end{equation}
\begin{equation}
    \fl 2 \tilde{\sigma}^{2} - \frac{2}{3}\tilde{\Theta}^{2} - \mathcal{Q} + 2\frac{\mathcal{Q} +\beta  \mathcal{Q}^{n+\frac{1}{2}}}{1 + 2 \beta  n \mathcal{Q}^{n-\frac{1}{2}}} + \frac{ 4 \sqrt{\mathcal{Q}} }{\alpha  \left(1 + 2 \beta  n Q^{n-\frac{1}{2}}\right)}\rho = 0,
\end{equation}
\begin{equation}\label{eq:3Ricci_Bianchi_alphasqrtQ}
    \fl \mathring{\tilde{\sigma}} + \tilde{\Theta} \tilde{\sigma} - \frac{1}{2} \frac{\mathring{\mathcal{Q}}}{\mathcal{Q}}\frac{1 - 4 \beta  (n-1) n \mathcal{Q}^{n-\frac{1}{2}}}{1 +  2 \beta  n \mathcal{Q}^{n-\frac{1}{2}}}\tilde{\sigma} = 0,
\end{equation}
\begin{equation}
    \fl \mathring{\rho} + \tilde{\Theta} \left( 1 + w \right)\rho = 0.
\end{equation}
\begin{table}
\centering
\renewcommand{\arraystretch}{1.5}
\caption{The stability of the fixed points and evolution of $l$, $\tilde{\sigma}$, and $\rho$ for $f(\mathcal{Q})=\alpha\left(\sqrt{\mathcal{Q}} + \beta \mathcal{Q}^{n}\right)$ and $\pi_{ij}= 0$. The parameters $\tau_{0}$, $l_{0}$, $\sigma_{0}$, $\sigma_{1}$, and $\rho_{0}$ are constants of integration.}
\begin{tabular}{ lcccccc }
\toprule
 & \multicolumn{6}{c}{\texorpdfstring{$0 \leq w \leq 1$}{}} \\
\cmidrule{2-7}
Point & \multicolumn{2}{c}{Attractor} & \multicolumn{2}{c}{Repeller} & \multicolumn{2}{c}{Saddle} \\ 
\midrule
$P_{1}$ &  \multicolumn{2}{c}{$n>\frac{1}{2}$} & \multicolumn{2}{c}{} & \multicolumn{2}{c}{} \\
$P_{2}$ & \multicolumn{2}{c}{} & \multicolumn{2}{c}{} &  \multicolumn{2}{c}{$n>\frac{1}{2}$}\\
\toprule
 &  \multicolumn{2}{c}{Average length} &  \multicolumn{2}{c}{Shear} & \multicolumn{2}{c}{Energy density}\\
\midrule
$P_{1}$ & \multicolumn{2}{c}{$l = l_{0} \left( \tau - \tau_{0} \right)^{\frac{2 n}{3 (1+w)}}$} & \multicolumn{2}{c}{$\tilde{\sigma}  = \sigma_{0} = 0$} & \multicolumn{2}{c}{$\rho = \rho_{0} = 0$}\\
$P_{2}$ & \multicolumn{2}{c}{$l = l_{0} \left( \tau - \tau_{0} \right)^{\frac{2 n}{3 (1 + 2n + w)}}$} & \multicolumn{2}{c}{$\sigma = \sigma_{0} + \sigma_{1} \left( \tau - \tau_{0} \right)^{-1}$} & \multicolumn{2}{c}{$\rho = \rho_{0} = 0$}\\
\bottomrule
\end{tabular}
\label{table_3}
\end{table}
By defining the dynamical variables,
\begin{equation}\label{eq:dynamical_variables_3}
    \Sigma^{2} = 3 \frac{\tilde{\sigma}^{2}}{\tilde{\Theta}^{2}}, \qquad 
    \mathcal{B} = \beta \tilde{\Theta}^{2n-1},
    \qquad
    \Omega^{2} = 3 \frac{1}{\alpha}\frac{1}{\tilde{\Theta}}\rho,
\end{equation}
the reduced system of dynamical equations is
\begin{eqnarray}
    \fl \: \: \Omega &=& \left(\frac{3}{2}\right)^{\frac{1-n}{2}}\sqrt{\left(2n - 1\right){\mathcal B}} \left(1 - \Sigma ^2\right)^{\frac{n}{2}}, \\
    \fl \frac{\hbox{d}\Sigma}{\hbox{d}\mathcal{T}} &=& -\frac{3 \Sigma  \left(1-\Sigma ^2\right) }{ 3^{n}\sqrt{2}+2^{n+1}\sqrt{3} n \mathcal{B} \left(1-\Sigma^{2}\right)^{n-\frac{1}{2}} }\left[ 3^{n}\sqrt{2} +\right. \nonumber\\
    \fl &&\left.+ 2^{n}\sqrt{3} \mathcal{B} (1 + w -2 n w)\left(1-\Sigma ^{2}\right)^{n-\frac{1}{2}} \right],\label{eq:dynamical_equation_Sigma_prime_3}\\
    \fl \frac{\hbox{d}\mathcal{B}}{\hbox{d}\mathcal{T}} &=& \frac{3 \left(1 - 2n\right) \mathcal{B}}{ n \left[3^{n}\sqrt{2} + 2^{n+1}\sqrt{3} n \mathcal{B} \left(1-\Sigma^{2}\right)^{n-\frac{1}{2}}\right]}\Big\{ 3^{n} \sqrt{2} n \Sigma^{2} +\nonumber\\
    \fl &&+ \frac{3^{n}}{\sqrt{2}}\left(1+w\right) + 2^{n} \sqrt{3} n \mathcal{B} \left[1 + w + \left(1 + w - 2nw\right) \Sigma^{2}\right] \left(1-\Sigma ^2\right)^{n-\frac{1}{2}}\Big\}.\label{eq:dynamical_equation_B_prime_3}
\end{eqnarray}
We assume $\mathcal{B} \geq 0$, $n \geq 1/2$, and $0 \leq \Sigma \leq 1$, so that $\Omega$ is non-negative.

The invariant submanifold $\Sigma = 0$ represents isotropic universes, whereas $\Sigma = 1$ anisotropic ones, and $\mathcal{B} = 0$, similarly to the previous section, a surface where the Lagrangian function is $f(\mathcal{Q}) = \alpha \sqrt{\mathcal{Q}}$, when $\beta$ is negligible, or the expansion rate $\tilde{\Theta}$ is zero. 

The critical points are
\begin{eqnarray}
    P_{1} &=& \lbrace \Sigma = 0,\: \mathcal{B} = 0,\: \Omega = 0 \rbrace ,\\
    P_{2} &=& \lbrace \Sigma = 1,\: \mathcal{B} = 0,\: \Omega = 0 \rbrace.
\end{eqnarray}
Both critical points have $\mathcal{B}$ and $\Omega$ equal to zero, and they are distinguished by the presence or absence of the shear $\Sigma$. The stability of the system and the approximate solutions are summarized in Table \ref{table_3}.

\begin{figure}[t]
\centering
\includegraphics[width=0.49\textwidth]{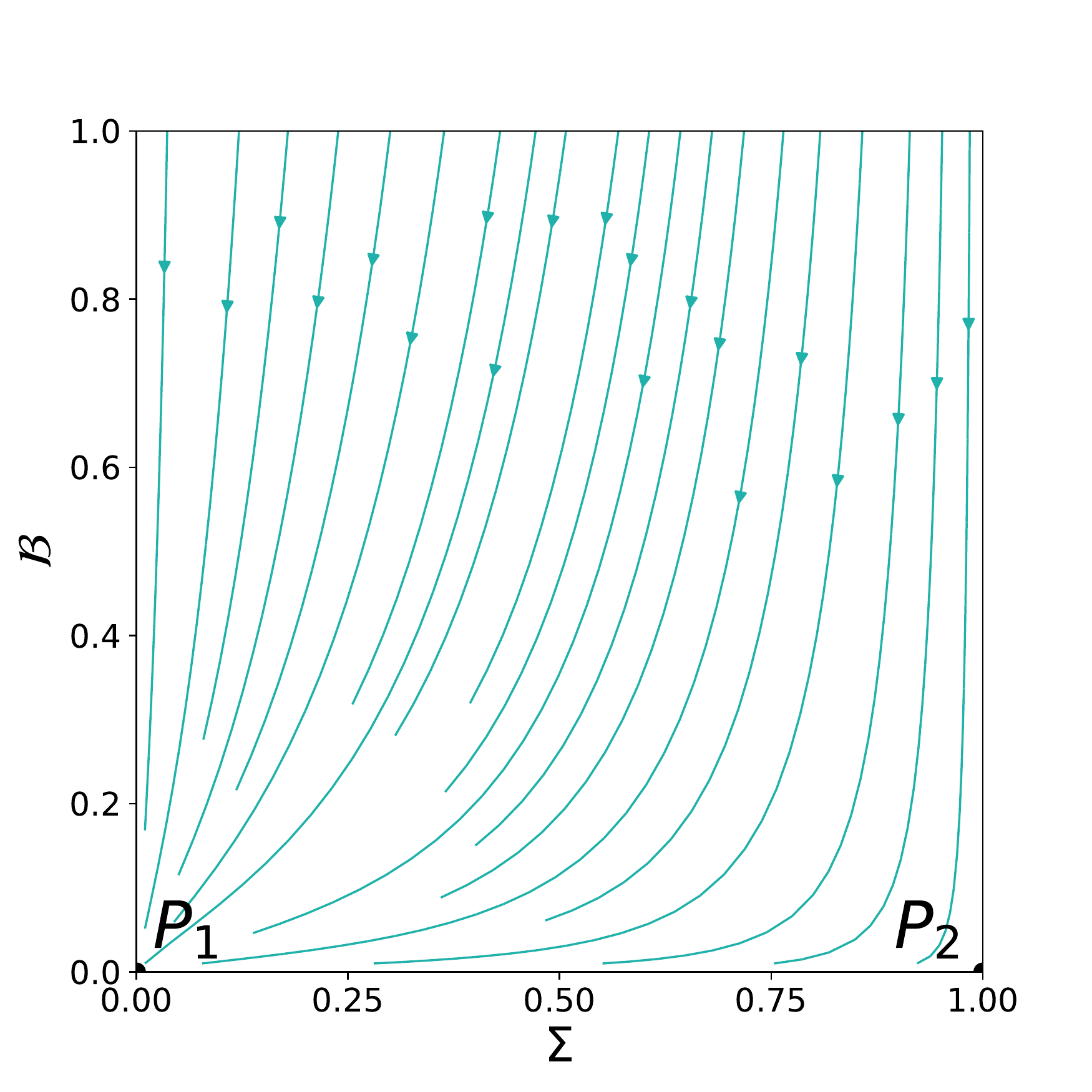}
\caption{Phase space portrait of the system \eref{eq:dynamical_equation_Sigma_prime_3}-\eref{eq:dynamical_equation_B_prime_3} for $w=0$ and $n=3$.}
\label{fig:example3_1}
\end{figure}

A representation of the stability is given in Figure \ref{fig:example3_1}. We notice that all the orbits converge in $P_{1}$, which is a global attractor. Hence, in this theory the universe always becomes isotropic.

\subsection{\texorpdfstring{$f(\mathcal{Q})$}{} as Lambert function}\label{sec:expQ}
For this last example we consider the function,
\begin{equation}
    f(\mathcal{Q}) = \mathcal{Q} \: e^{\alpha \mathcal{Q}},
\end{equation}
where $\alpha$ is a dimensional constant, and the anisotropic pressure $\pi_{ij}$ is zero.

The cosmological equations are,
\begin{equation}\label{eq:raychaudhuri_equation_bianchi_expQ}
    \fl \mathring{\tilde{\Theta}} + \frac{1}{3}\tilde{\Theta}^{2} + 2 \tilde{\sigma}^{2} + \frac{\alpha \mathcal{Q}^{2}}{2 \left(1 + \alpha  \mathcal{Q} \right)} + \frac{\alpha   (2 + \alpha  \mathcal{Q})}{1 + \alpha  \mathcal{Q}} \mathring{\mathcal{Q}} \tilde{\Theta}  + \frac{ e^{-\alpha \mathcal{Q}}}{2\left(1 + \alpha  \mathcal{Q}\right)}\left(1 + 3 w\right)\rho = 0,
\end{equation}
\begin{equation}
    \fl 2 \tilde{\sigma}^{2} - \frac{2}{3}\tilde{\Theta}^{2} - \mathcal{Q} + \frac{\mathcal{Q}}{1 + \alpha \mathcal{Q}} + \frac{2 \: e^{-\alpha \mathcal{Q}}}{1 + \alpha \mathcal{Q}} \rho = 0,
\end{equation}
\begin{equation}
    \fl \mathring{\tilde{\sigma}} + \tilde{\Theta} \tilde{\sigma} + \frac{\alpha \left(2 + \alpha \mathcal{Q}\right)}{1 + \alpha \mathcal{Q}} \mathring{\mathcal{Q}} \sigma = 0,
\end{equation}
\begin{equation}\label{eq:energy_momentum_conservation_bianchi_exp}
    \fl \mathring{\rho} + \tilde{\Theta} \left( 1 + w \right)\rho = 0.
\end{equation}
The introduction of the following dynamical variables,
\begin{equation}\label{eq:dynamical_variables_4}
    \Sigma^{2} = 3 \frac{\tilde{\sigma}^{2}}{\tilde{\Theta}^{2}}, \qquad 
    \mathcal{A} = \alpha \tilde{\Theta}^{2} ,
    \qquad
    \Omega^{2} = 3 \frac{1}{\tilde{\Theta}^{2}}\rho,
\end{equation}
leads to the equation for $\Omega$,
\begin{equation}
    \Omega = \sqrt{\left(1 - \Sigma^{2}\right)\left[1 + \frac{4}{3} \mathcal{A} \left(1-\Sigma^{2}\right)\right]} \: e^{\frac{1}{3} \mathcal{A} \left(1 - \Sigma^{2}\right)}
\end{equation}
and the system of two differential equations, 
\begin{eqnarray}
    \frac{\hbox{d}\Sigma}{\hbox{d}\mathcal{T}} &=& -\frac{3 \Sigma  \left(1-\Sigma^{2}\right) \big\{ 3 - w \left[3 + 4 \mathcal{A} \left(1-\Sigma^{2}\right) \right]\big\}}{2 \left[3 + 2 \mathcal{A} \left(1-\Sigma^{2}\right)\right]},\label{eq:dynamical_equation_Sigma_prime_4}\\
    \frac{\hbox{d}\mathcal{A}}{\hbox{d}\mathcal{T}} &=&  6 w \mathcal{A}\: \Sigma^{2} - \frac{9}{2} \left(1 + w\right)\mathcal{\mathcal{A}}\bigg\{\frac{2  \Sigma^{2}}{3 + 2 \mathcal{A} \left(1 - \Sigma^{2}\right)} +\nonumber\\ 
    && + \frac{2 \left[3 + 4 \mathcal{A} \left(1-\Sigma^{2}\right) \right]}{9 + 2 \mathcal{A} \left(1-\Sigma^{2}\right) \left[15 + 4 \mathcal{A} \left(1 - \Sigma^{2}\right) \right]}\bigg\}.\label{eq:dynamical_equation_A_prime_4}
\end{eqnarray}
To guarantee that $\rho>0$, we need to impose $\Omega \geq 0$, which in turn implies the conditions,
\begin{equation}\label{eq:condition_Omega_4_1}
    \mathcal{A} \leq -\frac{3}{4} \quad {\rm and} \quad  \frac{1}{2} \sqrt{\frac{3 + 4 \mathcal{A}}{\mathcal{A}}}\leq \Sigma \leq 1 
\end{equation}
or 
\begin{equation}\label{eq:condition_Omega_4_2}
    \mathcal{A} > -\frac{3}{4} \quad {\rm and} \quad 0\leq \Sigma \leq 1.
\end{equation}
We identify the invariant submanifolds $\Sigma = 0$, $\Sigma = 1$, and $\mathcal{A} = 0$. The first two outline isotropic and anisotropic universes, respectively; $\mathcal{A} = 0$ is the surface where the theory reduces to $f(\mathcal{Q})=\mathcal{Q}$, or to a cosmology where $\tilde{\Theta}=0$.

In the range given by Eqs. \eref{eq:condition_Omega_4_1} and \eref{eq:condition_Omega_4_2}, the critical points are,
\begin{eqnarray}
    P_{1} &=& \lbrace \Sigma = 0,\: \mathcal{A} = 0,\: \Omega = 1 \rbrace, \\
    P_{2} &=& \lbrace \Sigma = 1,\: \mathcal{A} = 0,\: \Omega = 0 \rbrace, \\
    P_{3} &=& \bigg\{ \Sigma = 0,\: \mathcal{A} = - \frac{3}{4},\: \Omega = 0 \bigg\}.
\end{eqnarray}
Moreover, for $w=1$ and $\mathcal{A}=0$, the system of Eqs. \eref{eq:dynamical_equation_Sigma_prime_4} and \eref{eq:dynamical_equation_A_prime_4} admits the solution,
\begin{equation}
    P_{4} = \Bigg\{ \Sigma = \Sigma^{*},\: \mathcal{A} = 0, \: \Omega = \sqrt{1 - \left( \Sigma^{*}\right)^{2}} \Bigg\},
\end{equation}
where $\Sigma^{*}$ is an arbitrary constant.

\begin{table}[t]
\centering
\renewcommand{\arraystretch}{1.5}
\caption{The stability of the fixed points and evolution of $l$, $\tilde{\sigma}$, and $\rho$ for $f(\mathcal{Q})=\mathcal{Q}e^{\alpha\mathcal{Q}}$ and $\pi_{ij}= 0$. The parameters $\tau_{0}$, $l_{0}$, $\sigma_{0}$, $\rho_{0}$ and $\rho_{1}$ are constants of integration.}
\begin{tabular}{ lcccccc  }
\toprule
Point & \multicolumn{2}{c}{Attractor} & \multicolumn{2}{c}{Repeller} & \multicolumn{2}{c}{Saddle} \\ 
\midrule
$P_{1}$ & \multicolumn{2}{c}{$0 \leq w < 1$} & \multicolumn{2}{c}{} & \multicolumn{2}{c}{} \\
$P_{2}$ & \multicolumn{2}{c}{$0 \leq w \leq 1$} & \multicolumn{2}{c}{} & \multicolumn{2}{c}{} \\
$P_{3}$ & \multicolumn{2}{c}{} & \multicolumn{2}{c}{} & \multicolumn{2}{c}{$0 \leq w < 1$} \\
$P_{4}$ & \multicolumn{2}{c}{$w=1$} & \multicolumn{2}{c}{} & \multicolumn{2}{c}{} \\
\toprule
 &  \multicolumn{2}{c}{Average length} &  \multicolumn{2}{c}{Shear} & \multicolumn{2}{c}{Energy density}\\
\midrule
$P_{1}$ & \multicolumn{2}{c}{$l(\tau) = l_{0} \left( \tau - \tau_{0} \right)^{\frac{2}{3 (1+w)}}$} & \multicolumn{2}{c}{$\tilde{\sigma} = \sigma_{0} = 0$} & \multicolumn{2}{c}{$\rho (\tau) = \rho_{0} + \frac{\rho_{1}}{\left( \tau - \tau_{0} \right)^{2}}$}\\
$P_{2}$ & \multicolumn{2}{c}{$l(\tau) = l_{0} e^{\frac{\tau}{\tau_{0}}}$} & \multicolumn{2}{c}{$\tilde{\sigma} = \sigma_{0} = 0$} & \multicolumn{2}{c}{$\rho(\tau) = \rho_{0} = 0$}\\
$P_{3}$ & \multicolumn{2}{c}{$l(\tau) = l_{0} \sqrt[3]{3\left( \tau - \tau_{0}\right)}$} & \multicolumn{2}{c}{$\tilde{\sigma} (\tau) = \sigma_{0} + \frac{1}{\sqrt{3} (\tau - \tau_{0})}$} & \multicolumn{2}{c}{$\rho(\tau) = \rho_{0} = 0$}\\
$P_{4}$ & \multicolumn{2}{c}{$l(\tau) = l_{0} \sqrt[3]{3\left( \tau - \tau_{0}\right)}$} & \multicolumn{2}{c}{$\tilde{\sigma} (\tau) = \sigma_{0} + \frac{\Sigma^{*}}{\sqrt{3} (\tau - \tau_{0})}$} & \multicolumn{2}{c}{$\rho (\tau) = \rho_{0} + \frac{1 - \Sigma^{*}{}^{2}}{3 \left(\tau - \tau_{0}\right)^{2}}$}\\
\bottomrule
\end{tabular}
\label{table_4}
\end{table}

The results of the stability analysis near the critical points and the approximate solutions are outlined in Table \ref{table_4}.

\begin{figure}[t]
\centering
    \begin{subfigure}[ht]{0.49\textwidth}
        \includegraphics[width=\linewidth]{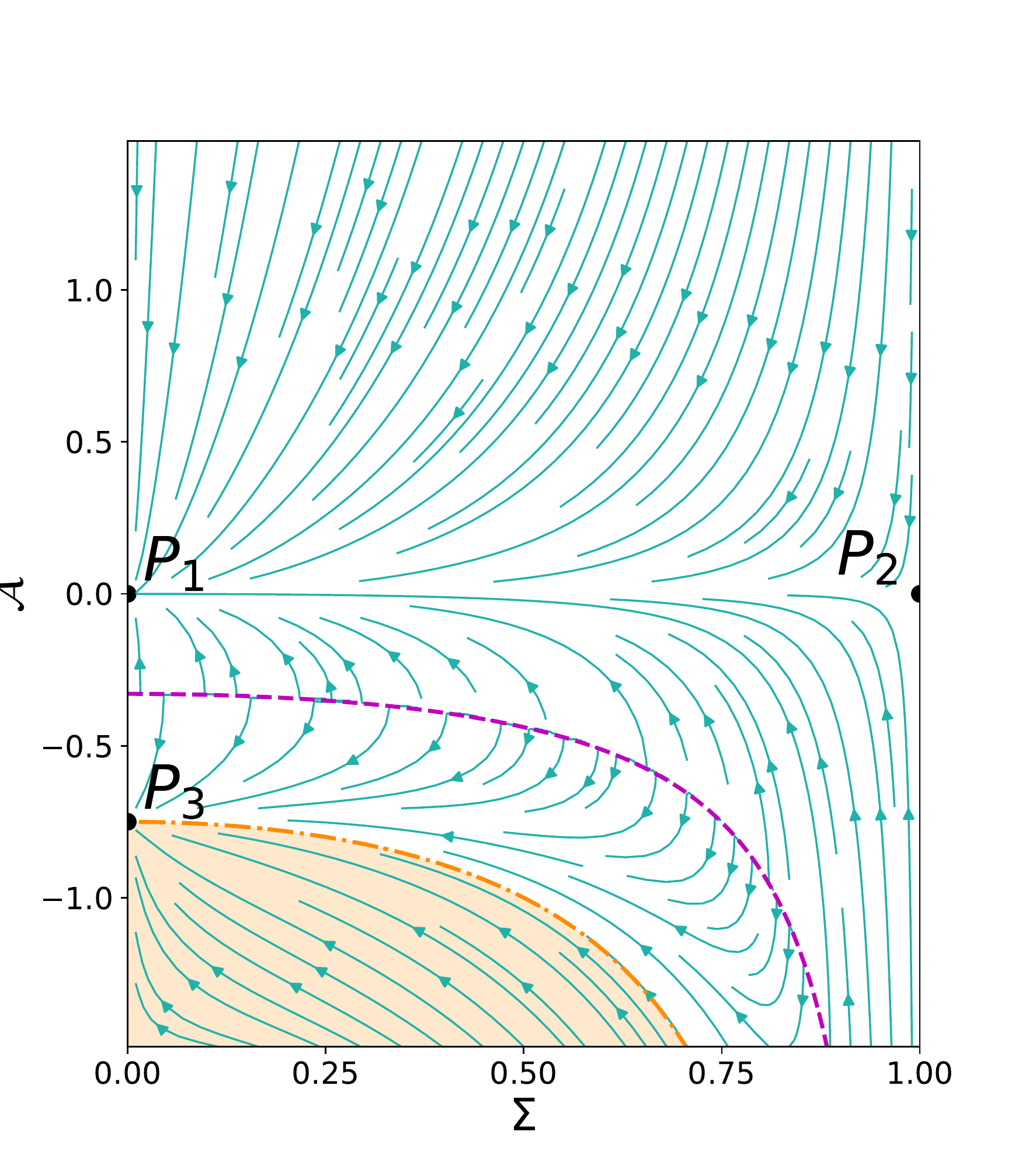}
        \caption{}
        \label{fig:Phase_space_alpha_expQ_1}
    \end{subfigure}
    \begin{subfigure}[ht]{0.49\textwidth}
        \includegraphics[width=\linewidth]{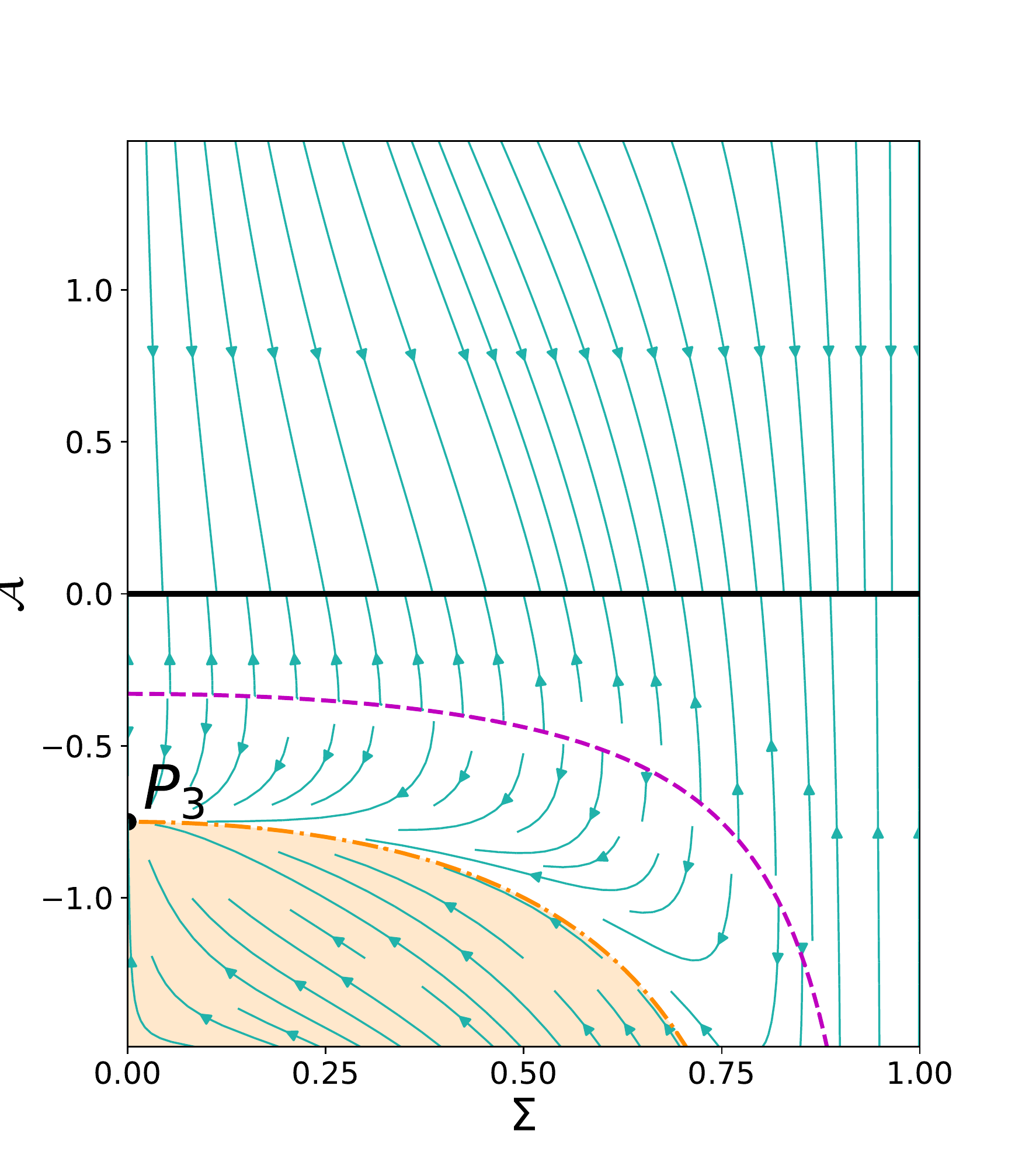} 
        \caption{}
        \label{fig:Phase_space_alpha_expQ_2}
    \end{subfigure}
     \caption{Phase space portrait of the system \eref{eq:dynamical_equation_Sigma_prime_4}-\eref{eq:dynamical_equation_A_prime_4} for (a) $w=0$, and (b) $w=1$. Shaded areas are non-physical regions for the phase space.}
    \label{fig:Phase_space_alpha_expQ}
\end{figure}

The phase space of the Eqs. \eref{eq:dynamical_equation_Sigma_prime_4}, and \eref{eq:dynamical_equation_A_prime_4} is represented in Figure \ref{fig:Phase_space_alpha_expQ}. In Figure \ref{fig:Phase_space_alpha_expQ_1}, $P_{1}$ and $P_{3}$ are attractors, and $P_{2}$ is a saddle point. In Figure \ref{fig:Phase_space_alpha_expQ_2}, in addition to the point $P_{3}$, all the space identified by $\mathcal{A}=0$, i.e. the central heavy line in the figure, is an attractor. In both figures, the phase space is divided into three regions by two curves. The dash-dotted line indicates the curve
\begin{equation}
 3 + 4 \mathcal{A}\left(1-\Sigma^{2}\right)=0,
\end{equation}
determining the lower boundary for which, by Eq. \eref{eq:condition_Omega_4_1}, $\Omega$ is positive. Therefore, the phase space is not physical below this line and in the figures correspond to the shaded area. Instead, the dashed curve represents one of the denominators of Eq. \eref{eq:dynamical_equation_A_prime_4},
\begin{equation}
 9 + 2 \mathcal{A} \left(1-\Sigma^{2}\right) \left[15 + 4 \mathcal{A} \left(1 - \Sigma^{2}\right) \right]=0,
\end{equation}
the other denominator of Eq. \eref{eq:dynamical_equation_A_prime_4} is irrelevant as it lays below the dash-dotted curve.
 
The presence of the sectors delimited by the dashed and dash-dotted curves is an essential difference from the other examples discussed above. In Sec. \ref{sec:alphaQ_anisotropic_pressure} we have analyzed different behaviors of the orbits according to the positivity or negativity of the constants related to the dynamical parameter . Here, however, for $\alpha<0$ there are different attractors, depending on whether an orbit is above or below the divergence line. Therefore, the final state of cosmology depends crucially on the initial conditions. For example, in the case $w=1$, the orbits below the divergence line describe universes which tends toward isotropy, whereas orbits above it tend to a finite value of  $\Sigma$, i.e. the universe approaches an anisotropic state.


\section{Discussion and conclusions}\label{sec:conclusions}
We investigated the dynamics of Bianchi type-I cosmologies within the framework of $f(\mathcal{Q})$ gravity using a combination of the $1+3$ covariant formalism and the dynamical systems approach.

The $1+3$ formalism allowed us to obtain a very clear and detailed description of the geometric and dynamic properties of $f(\mathcal{Q})$ cosmologies. In particular, we were able to characterize the effect of nonmetricity on the autoparallel motion of the observers and to obtain  cosmological equations which are independent of any specific coordinate system. In addition, the $1+3$  decomposition made it possible to single out the different contributions of the of the nonmetricity tensor $Q_{kij}$, making more explicit the effect of nonmetricity on the kinematic quantities. We proved that in Bianchi type-I metric the decomposition of the tensor $Q_{kij}$ involves only the scalar and traceless symmetric tensors which affect the expansion rate $\Theta$ and shear $\sigma$. 

One of the main difficulties of applying the $1+3$ formalism to nonmetric theories of gravity is that in general one cannot assume proper time to be an affine parameter along the timelike congruence. However, in the case of Bianchi type-I cosmologies this problem can be overcome, thus obtaining complete equivalence between the affine parameter of the world lines and the proper time of the observers associated with the congruence. This aspect is crucial, allowing the introduction of an unambiguous cosmic time and then the definition of a cosmic history.

After writing the cosmological equations in the $1+3$ framework, we separated the contributions due to Levi-Civita from the nonmetricity terms, in order to better understand the differences between GR and $f(\mathcal{Q})$ gravity. As it happens in many other extensions of GR, we were able to describe in a complete way the additional terms that nonmetricity induces in the gravitational field equations as contributions due to an effective energy-momentum tensor. This formulation allowed an immediate application of the DSA. Although semi-quantitative, a phase space analysis of the $f(\mathcal{Q})$ cosmological models allows us to derive several interesting and general features. We considered here four applications, involving different functions $f(\mathcal{Q})$ and thermodynamical properties of the sources.

In the first application, the function $f(\mathcal{Q})$ was a power law (Sec. \ref{sec:alphaQ}), which was chosen because of its simplicity and because it is commonly used in literature. We obtained a one-dimensional dynamical system which was solvable analytically. We compared the results with those of the paper \cite{Esposito:2021ect}, exhibiting a perfect match when the universe, filled with dust, is initially anisotropic and then isotropizes. This is not surprising as the phase space contains all cosmological solutions, and thus it must include the one reconstructed in \cite{Esposito:2021ect}. 

We also analyzed a cosmology with the same power law action, but in the presence of an anisotropic pressure which we assumed proportional to the shear (Sec. \ref{sec:alphaQ_anisotropic_pressure}). In this scenario, an isotropic universe is seen to have a transition phase associated with a saddle point, from which the orbits either diverge completely from it or return to the anisotropic state from which they started. This behavior suggests a universe with a ``cyclic'' evolution, in which after a phase of isotropy, anisotropies start to grow again.

In \cite{Esposito:2021ect} it was found that the reconstructed forms of $f(\mathcal{Q})$ always have a $\sqrt{\mathcal{Q}}$ term which plays a role similar to an integration constant. As another application (Sec. \ref{sec:sqrtQ}), we investigated the effect of this term when it is added in the functions $f(\mathcal{Q})$ used in the previous two examples. Our analysis showed that the main effect of this additional term is, as expected, constraining the sign of the nonmetricity scalar $\mathcal{Q}$, which in turn excludes some possible cosmic histories (the ones for $\Sigma>1$). In the cases we considered, the additional term forces all cosmologies to become isotropic in the future.

As a final example, we attempted the evaluation of the effects due to a gravitational action consisting of an infinite series of power law terms. Such effects can be evaluated considering a function $f(\mathcal{Q})$ as the Lambert function (Sec. \ref{sec:expQ}). In this case, the phase space differs considerably from the ones of the previous examples. The most important difference turned out to be the appearance of separate regions of the phase space. The presence of these regions shows that the cosmology will have different behaviors and different final attractors depending on the initial conditions.  

In all the examples we considered, some areas of the phase space needed to be excluded. We saw that these forbidden regions can appear for different reasons. For instance, in Secs \ref{sec:alphaQ} and \ref{sec:alphaQ_anisotropic_pressure},  the chosen dynamical variables and the request to have a matter with physical thermodynamical quantities implied the exclusion of the line $\Sigma = 1$. In other cases, like the ones given in Sec. \ref{sec:sqrtQ} and \ref{sec:expQ}, the limitations were related to the nature of the function $f(\mathcal{Q})$. For example, the condition $\Sigma \neq 1$ is connected to the fact that the function $f(\mathcal{Q})$ might take, along an orbit, values that change dramatically the structure of the gravitational field equations, giving rise to singularities or degeneracies.

We conclude by remarking that the DSA, especially combined with the $1+3$ covariant approach, showed yet again a great potential in clarifying the physics of cosmological models. In particular, $f(\mathcal{Q})$  cosmology exhibits a behavior of the anisotropy which is much richer than the one of GR, and this constitutes an important element in the search for experimental constraints of these models. Moreover, a deeper understanding of the differences between $f(\mathcal{Q})$ gravity and other extensions or modifications of GR will be certainly a challenge for future investigations. 


\appendix

\section{\texorpdfstring{$f(\mathcal{Q})$}{} field equations}\label{appendix:field_equations_derivation}
Due to the condition $R^{h}{}_{kij}=0$ and after contracting the first and third index, we can rewrite Eq. \eref{eq:def_riemann_1} in the form,
\begin{equation}
    \tilde{R}_{ij} =  -\tilde{\nabla}_{h}N_{ij}{}^{h}+\tilde{\nabla}_{j}N_{hi}{}^{h} - N_{hk}{}^{h}N_{ij}{}^{k} + N_{jk}{}^{h}N_{hi}{}^{k}.
\end{equation}
Moreover, inserting the identities,
\begin{equation}
\eqalign{
    \fl \frac{2}{\sqrt{-g}}\nabla_{h} \left( \sqrt{-g} f' P^{h}{}_{ij} \right) =&  2f'\tilde{\nabla}_{h}P^{h}{}_{ij} - 2f' N_{hi}{}^{k}P^{h}{}_{kj} +\\
    &- 2f'N_{hj}{}^{k}P^{h}{}_{ki} + 2 f'' \partial_{h} \mathcal{Q}P^{h}{}_{ij},
    }    
\end{equation}
and
\begin{equation}
    \fl N_{jk}{}^{h}N_{hi}{}^{k} - N_{hk}{}^{h}N_{ij}{}^{k} = - 2 N_{hi}{}^{k}P^{h}{}_{kj} - 2N_{hj}{}^{k}P^{h}{}_{ki} + P_{iab}Q_{j}{}^{ab} - 2 Q^{ab}{}_{i}P_{abj},
\end{equation}
into Eq. \eref{eq:metric_equation}, we get
\begin{equation}\label{eqappendix:field_equation_1}
    \fl 2 f' \tilde{\nabla}_{h}P^{h}{}_{ij} + \frac{1}{2}g_{ij}f + f' \tilde{R}_{ij} +2f''\partial_{h}\mathcal{Q}P^{h}{}_{ij} + f' \left( \tilde{\nabla}_{h}N_{ij}{}^{h}  - \tilde{\nabla}_{j}{N}_{hi}{}^{h}\right)  = \Psi_{ij}.
\end{equation}
From the trace of \eref{eqappendix:field_equation_1} we have,
\begin{equation}
    \fl \tilde{R} = \frac{1}{f'}\Psi - 2 \frac{f}{f'} - 2\tilde{\nabla}_{h}P^{hi}{}_{i} -2 \tilde{\nabla}_{[h}N_{i]}{}^{ih} - 2 \frac{f''}{f'}P^{h}{}{ij}\partial_{h}\mathcal{Q}.
\end{equation}
Subtracting $\frac{1}{2}g_{ij}\tilde{R}$ from \eref{eqappendix:field_equation_1}, we have also,
\begin{equation}
\eqalign{
    \fl \tilde{R}_{ij} - \frac{1}{2}g_{ij}\tilde{R} =& \frac{1}{f'}\left(\Psi_{ij} - \frac{1}{2}g_{ij}\Psi \right) + \frac{1}{2}g_{ij}\frac{f}{f'} - g_{ij}\tilde{\nabla}_{h}P^{hk}{}_{k} + \\ 
    \fl &- \tilde{\nabla}_{h}N_{ij}{}^{h} + \tilde{\nabla}_{j}N_{hi}{}^{h} 
    - 2 \frac{f''}{f'}\left( P^{h}{}_{ij} - \frac{1}{2}g_{ij}P^{hk}{}_{k} \right) +\\
    \fl &- 2\tilde{\nabla}_{h}P^{h}{}_{ij} + \frac{1}{2}g_{ij}\left( \tilde{\nabla}_{h}N_{k}{}^{kh} - \tilde{\nabla}_{k}N_{h}{}^{kh} \right).
    }  
\end{equation}
Moreover, from Eq. \eref{eq:def_homothetic_tensor}, the further identity
\begin{equation}
\eqalign{
    \fl 0 = \frac{1}{2}\hat{R}_{ij} &= -\frac{1}{2}\tilde{\nabla}_{[i}\mathcal{Q}_{j]h}{}^{h} =\\
    \fl &= 2 \tilde{\nabla}_{h}P^{h}{}_{ij} + \tilde{\nabla}_{h}N_{ij}{}^{h} - \tilde{\nabla}_{j}N_{hi}{}^{h} - \frac{1}{2}g_{ij} \left( \tilde{\nabla}_{h}N_{k}{}^{kh} - \tilde{\nabla}_{h}N_{k}{}^{hk}\right),
    }    
\end{equation}
implies
\begin{equation}\label{A.1}
\eqalign{
    \fl \tilde{R}_{ij} - \frac{1}{2}g_{ij}\tilde{R} =& \frac{1}{f'}\left(\Psi_{ij} - \frac{1}{2}g_{ij}\Psi \right) + \frac{1}{2}g_{ij}\frac{f}{f'} +\\
    \fl &+ g_{ij}\tilde{\nabla}_{h}P^{hk}{}_{k} - 2 \frac{f''}{f'}\left( P^{h}{}_{ij} - \frac{1}{2}g_{ij}P^{hk}{}_{k} \right)\partial_{h}\mathcal{Q}. 
    }   
\end{equation}
Inserting  Eq. \eref{eq:ricci_scalar_Q} into Eq. \eref{A.1}, we obtain the final equation,
\begin{equation}
    \fl \tilde{R}_{ij} = \frac{1}{f'} \left( \Psi_{ij} - \frac{1}{2}g_{ij}\Psi \right) + \frac{1}{2}g_{ij}\left(\frac{f}{f'} - \mathcal{Q}\right) - 2 \frac{f''}{f'}\left( P^{h}{}_{ij}-\frac{1}{2}g_{ij}P^{hk}{}_{k} \right)\partial_{h}\mathcal{Q}.
\end{equation}


\section{Bianchi type-I model in coordinates}\label{appendix:bianchi_coordinates}
Bianchi type-I universes admit local coordinates in which the line element is expressed as,
\begin{equation}\label{eq:bianchi_metric}
    \hbox{d}s^2 = - \hbox{d}t^2 + a^2(t)\hbox{d}x^2 + b^2(t)\hbox{d}y^2 + c^2(t)\hbox{d}z^2,
\end{equation}
where $a(t)$, $b(t)$, and $c(t)$ are the scale factors of each spatial direction. On the other hand, in $f(\mathcal{Q)}$ gravity flatness and torsionless conditions ensure the existence of local coordinates in which $\Gamma_{ij}{}^{h}=0$. This is the so-called coincident gauge. 

Then, we assume that the coordinates \eref{eq:bianchi_metric} are precisely those which realize the coincidence gauge. Under this assumption, defining the $4$-velocity as
\begin{equation}
    u^{i} = \left( \phi(t),\: 0,\: 0,\: 0\right),
\end{equation}
and the spatial volume
\begin{equation}
    \tau(t) = a(t)b(t)c(t),
\end{equation}
a straightforward calculation shows that the only non-zero projections of the nonmetricity tensor $Q_{hij}$ are,
\begin{equation}
    Q_{1} = 2 \phi \frac{\dot{\tau}}{\tau},
\end{equation}
\begin{equation}
    Q^{(0)}{}_{11} = \phi \left(2a\dot{a} - \frac{2}{3}a^{2}\,\frac{\dot{\tau}}{\tau}\right),
\end{equation}
\begin{equation}
    Q^{(0)}{}_{22} = \phi \left(2b\dot{b} - \frac{2}{3}b^{2}\,\frac{\dot{\tau}}{\tau}\right),
\end{equation}
\begin{equation}
    Q^{(0)}{}_{33} = \phi \left(2c\dot{c} - \frac{2}{3}c^{2}\,\frac{\dot{\tau}}{\tau}\right).
\end{equation}

Notice that the autoparallelism and the condition \eref{eq:condition_about_u} are automatically satisfied. $u^{i}\nabla_{i}u^{k}=0$ is commonly imposed, whereas $Q_{kij}u^{k}u^{i}u^{j} = Q_{0}$ is identically zero, as proved above. Therefore, we can use the proper time to define the $4$-velocity, which can be normalized as
\begin{equation}\label{eq:appendix_u_normalized}
    u^{i} = (1,\: 0,\: 0,\: 0).
\end{equation}
Equation \eref{eq:appendix_u_normalized} can also be derived using both the condition $u^{i}\nabla_{i}u^{k}=0$ and the coincident gauge,
\begin{equation}
    u^{i}\nabla_{i}u^{k} = u^{i}\partial_{i}u^{k} \quad \rightarrow \quad  \phi\:\partial_{0}\:\phi = 0,
\end{equation}
which implies $\phi= const.$

In such a circumstance, the nonmetricity scalar is given by,
\begin{equation}
    \mathcal{Q} = - \frac{1}{4}Q^{(0)}{}_{ij}Q^{(0)}{}^{ij} + \frac{1}{6}Q_{1}{}^{2} =  2 \left( \frac{\dot{a}\dot{b}}{ab} + \frac{\dot{a}\dot{c}}{ac} +\frac{\dot{b}\dot{c}}{bc} \right).
\end{equation}

\newpage

\bibliographystyle{iopart-num.bst}
\bibliography{main_v3}

\end{document}